\documentclass[11pt]{article}

\usepackage[final]{acl}

\usepackage{times}
\usepackage{latexsym}

\usepackage[T1]{fontenc}

\usepackage[utf8]{inputenc}

\usepackage{microtype}

\usepackage{inconsolata}
\usepackage{subcaption}
\usepackage{graphicx}
\usepackage{amsmath}
\usepackage{tcolorbox}
\tcbuselibrary{breakable}
\usepackage{listings}
\usepackage{url}
\usepackage{booktabs}
\usepackage{amssymb}
\usepackage{tcolorbox}
\tcbuselibrary{listings}
\usepackage{cleveref}
\usepackage{pifont} 
\newcommand{\cmark}{\ding{51}}
\newcommand{\xmark}{\ding{55}}
\usepackage{kotex} 
\usepackage{fvextra}
\usepackage{tabularx}
\usepackage{multirow}
\usepackage{enumitem}

\DefineVerbatimEnvironment{CodeBlock}{Verbatim}
{fontsize=\footnotesize, breaklines=true, breakanywhere=true, showspaces=false, showtabs=false, obeytabs=false, breaksymbolleft=, breaksymbolright=}

\newif\ifrevision
\revisiontrue

\definecolor{revcolor}{RGB}{0,0,0}
\newcommand{\rev}[1]{%
  \ifrevision
    {\color{revcolor}#1}%
  \else
    #1%
  \fi
}

\newif\ifcameraready
\camerareadytrue

\definecolor{cmcolor}{RGB}{0,0,0}
\newcommand{\cm}[1]{%
  \ifcameraready
    {\color{cmcolor}#1}%
  \else
    #1%
  \fi
}

\title{ExPerT: Personalizing LLM Responses to Users' Domain Expertise \\ via Query-Wise Semantic and \rev{Keystroke} Behavioral Cues}

\author{
 \textbf{Yeji Park\textsuperscript{1}},
 \textbf{Jiwon Tark\textsuperscript{2}},
 \textbf{Taesik Gong\textsuperscript{1}}
\\
\\
 \textsuperscript{1} UNIST,
 \textsuperscript{2} Korea University
\\
 \texttt{yejipark@unist.ac.kr, goneii@korea.ac.kr, taesik.gong@unist.ac.kr}
}

\newcommand{\system}{ExPerT}
\begin{document}

\maketitle

\begin{abstract}
Large language models (LLMs) are increasingly used by end users, yet existing personalization methods relying on static profiles or text-only signals fail to capture query-specific expertise variation. We present \system{}, a query-wise personalization framework that adapts LLM responses to users' query domain expertise by combining semantic and behavioral cues. \system{} consists of two key components:  (i) a semantic–behavioral expertise inference module that jointly interprets query text and keystroke dynamics via in-context LLM prompting, and (ii) an expertise-conditioned response generation that adapts the level of detail, terminology, and conceptual complexity. \rev{Our user study with 40 participants and 1270 queries demonstrated that \system{} reduced expertise inference error by 65.7\% compared to the strongest baseline (MAE $=$ 0.398 vs. 1.162) and improved response satisfaction by 17.52\% (from 3.71 to 4.36) on a 5‑point Likert scale.}

\end{abstract}

\begin{figure*}[t]
    \centering
    \includegraphics[width=\linewidth]{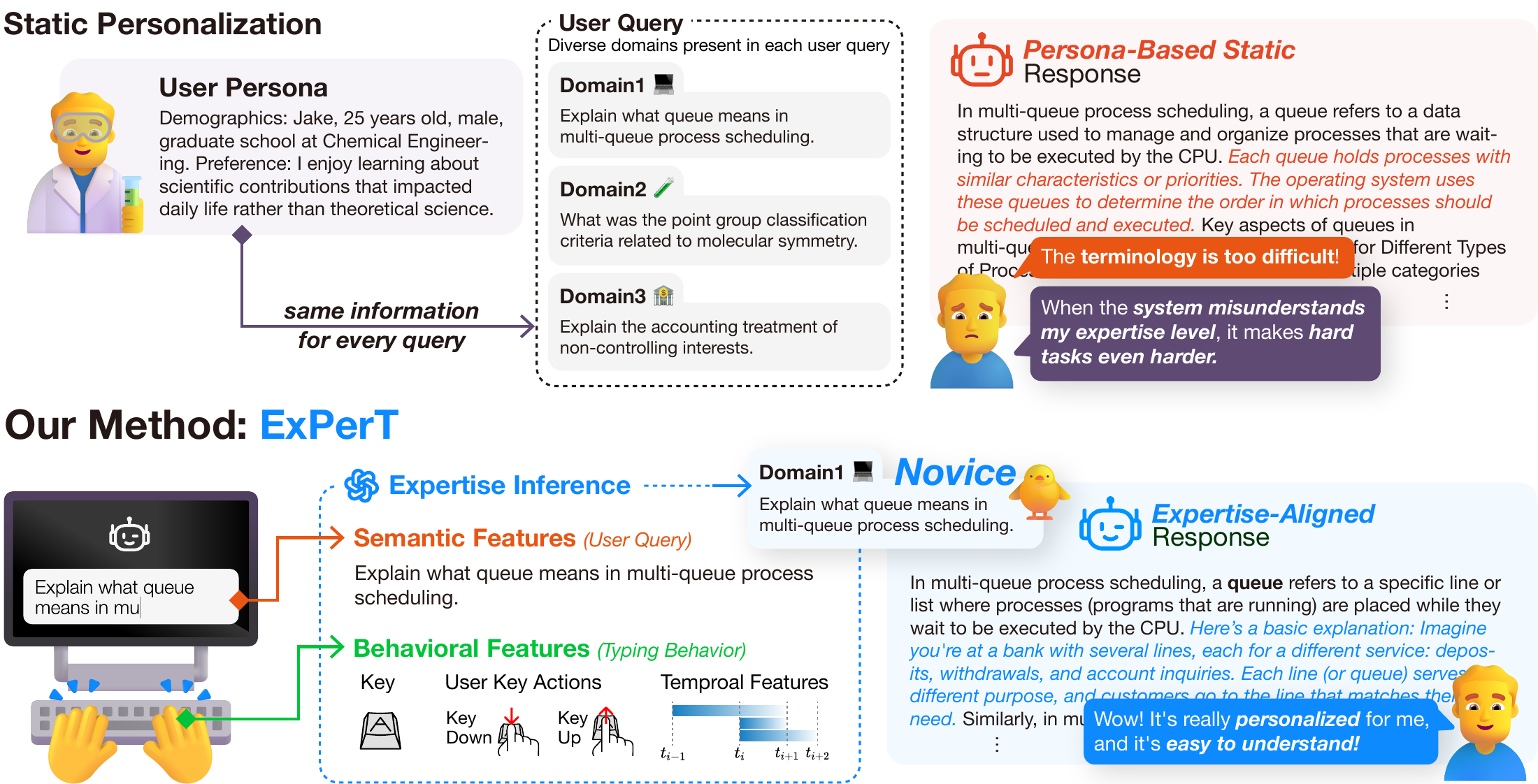}
    \caption{Comparison of static personalization (previous approaches) and \system{}'s query-wise, expertise-adaptive LLM responses. Static personalization applies the same fixed user profile to all queries. \system{} dynamically infers expertise from keystroke behavior for each query, enabling query domain expertise-aligned responses.}
    \label{fig:main_fig}
\end{figure*}

\section{Introduction}

Owing to recent advances and the widespread deployment of LLMs, they are now integrated into a broad range of daily tasks, including workflow planning~\cite{chan2023mango, lin2025seeking}, decision-making processes~\cite{eigner2024determinants, chiang2024enhancing}, and reasoning~\cite{wei2022chain, wang2023can}.
Consequently, users now anticipate responses that are not only accurate but also personalized to individual needs~\cite{wang2023cue}, goals~\cite{dang2025authoring}, and backgrounds~\cite{wang2024ai}.
To address these expectations, recent studies have focused on personalizing LLM responses by integrating user-specific text-based information, including personas~\cite{hu-collier-2024-quantifying, sun-etal-2025-persona}, preferences~\cite{zhao2025llms}, and long-term user histories~\cite{magister2024way}. 
However, these static profile-based approaches~\cite{hu-collier-2024-quantifying, sun-etal-2025-persona} often fail to capture per-query dynamics, where user context can shift depending on domain expertise, task demands, or intent~\cite{wang2024ai, cheng2024dialogues}.

To support query-wise dynamic personalization, we focus on user domain expertise---an essential factor that determines the appropriate level of detail, terminology, and conceptual complexity in LLM-generated responses~\cite{10.1145/3411764.3445648,zhang-etal-2024-knowledgeable,10.1145/3589334.3645404,tang2025dynamiccontextawarepromptrecommendation}. Notably, a user's expertise can vary significantly across queries, even within the same subject, making it a crucial signal for generating personalized responses~\cite{kiseleva2015impacttechnicaldomainexpertise, palta2025speakingrightlanguageimpact}.
Yet, users' domain expertise is difficult to infer from query text alone---short or ambiguous prompts may obscure actual expertise level~\cite{10.1145/3616855.3635845}; for instance, experts may simplify their question for efficiency, while novices may adopt technical terms without fully understanding them. These challenges motivate us to consider combining textual cues with behavioral signals to infer expertise robustly at the query level.
To this end, we propose \system{} (Expertise-aware Personalized Text generation), a query-wise response personalization framework that dynamically adapts LLM responses to a user's query domain expertise by leveraging both semantic and behavioral cues (Figure~\ref{fig:main_fig}). Specifically, \system{} introduces an expertise inference module that jointly interprets (i) query semantics, such as phrasing and domain-specific terminology, and (ii) keystroke dynamics reflecting fluency and confidence during query composition to estimate user expertise.
To achieve this, we design an input feature extraction pipeline that combines query text with keystroke-derived word-level features, capturing timing and fluency patterns, to provide input features for the LLM. The inferred expertise level is subsequently injected into our expertise-conditioned generation prompt that encodes explicit instructions. 

We conducted experiments to validate two key assumptions of this work: (i) user expertise can be accurately inferred from semantic and behavioral cues, and (ii) personalizing LLM responses based on user expertise improves response satisfaction. We recruited \rev{40 participants} from three domains (chemistry, computer science, and business) who interacted through a custom chatbot interface that recorded query text, keystroke dynamics, self-reported expertise, and satisfaction ratings, yielding a dataset of \rev{1270} annotated queries. Using this dataset, we first evaluated whether semantic-behavioral cues enable accurate query-level expertise inference. \rev{Our approach achieved the lowest mean absolute error (MAE) of 0.398, representing a 65.7\% reduction compared to IDL~\cite{cheng2024dialogues}, the strongest baseline (MAE $=$ 1.162).
We then examined whether expertise-conditioned responses improve satisfaction and observed a 17.52\% increase (from 3.71 to 4.36) on a 5‑point Likert scale.} 
These results demonstrate that \system{} not only infers situational expertise accurately but also generates responses that better align with users' knowledge, improving perceived quality.



We summarize our key contributions as follows:
\begin{itemize}[leftmargin=*, itemsep=0pt, topsep=2pt]
    \item We present \system{}, a query-wise personalization framework that dynamically adapts LLM responses to users' domain expertise, addressing the limitations of existing static personalization.
    
    \item We introduce a semantic-behavioral expertise inference framework that jointly interprets query semantics and keystroke dynamics to accurately estimate expertise level without requiring explicit user profiles or long-term interaction history.

    \cm{\item We present a system design for this problem that combines: (i) a keystroke-feature extraction pipeline that converts raw typing events into structured word-level behavioral summaries, (ii) a structured few-shot in-context prompting scheme for expertise inference, (iii) joint semantic and behavioral reasoning for expertise prediction, and (iv) a two-stage expertise-conditioned response generation framework.}
    
    \item 
    We demonstrate, through a user study across three domains, that expertise-conditioned response generation significantly improves response alignment and user satisfaction.
\end{itemize}

The preprocessing pipeline and an anonymized version of the dataset are available.\footnote{\url{https://github.com/UbiquitousAILab/ExPerT}}

\begin{figure*}[t]
    \captionsetup{labelfont={color=revcolor}, textfont={color=revcolor}}
    \centering
    \includegraphics[width=\linewidth]{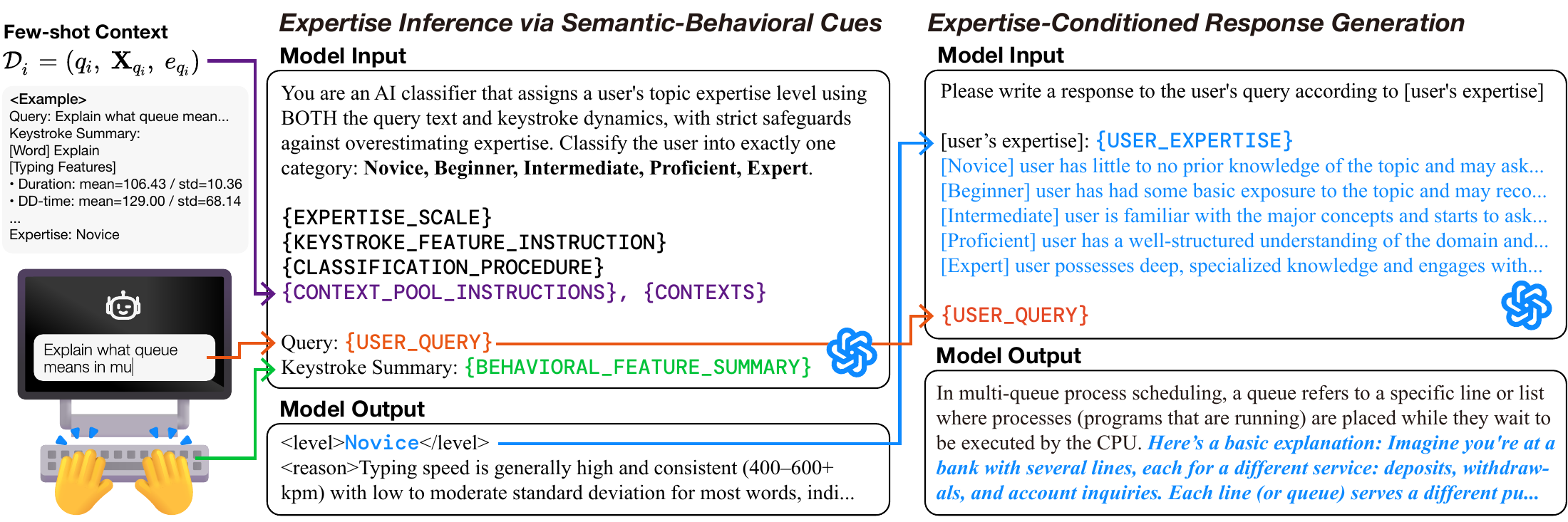}
    \caption{Overview of \system{}: (i) Given a user query and summary of behavioral data, \system{} infers user expertise from LLMs via in-context learning. (ii) The predicted expertise level is then used to generate a response personalized to the user's expertise level on the query.}
    \label{fig:method}
\end{figure*}

\section{Method}
We present \system{} (Expertise-aware Personalized Text generation), a query-wise adaptive response generation framework that personalizes LLM responses based on the user's domain expertise level. 
As shown in Figure~\ref{fig:method}, \system{} consists of two main modules: (i) an expertise inference module that estimates expertise from query text and keystroke dynamics via in-context learning (Section~\ref{sec:keystroke}), and (ii) a response generator that adjusts content and style using expertise-conditioned prompts (Section~\ref{sec:response}). Full prompt templates and examples are provided in Appendix~\ref{appendix:expertise_inference} and ~\ref{appendix:expertise-conditioned-generation}.


\subsection{Problem formulation}
\label{sec:problem}
Given a user query $q$, our objective is to generate a personalized response $r_q$ tailored to the user's domain expertise level, denoted as $e_q$.  This is achieved through a two-stage: (i) \textit{semantic-behavioral expertise inference}, where expertise is estimated from the user's semantic features derived from the query $q$ and behavioral features $\mathbf{X}_q$ such as a summary of keystroke dynamics, which are collectively represented as input $I_q$, and (ii) \textit{expertise-conditioned generation}, where the response is personalized based on the inferred expertise level. Formally:
\begin{align*}
\text{Input features:} \quad & I_q = \{q, \mathbf{X}_q\}, \\
\text{Inferred expertise level:} \quad & \hat{e}_q = f(I_q),\\
\text{Personalized response:} \quad & r_q = g\bigl(q, \hat{e}_q\bigr),
\end{align*}
where $f$ denotes the expertise inference function, and $g$ is the expertise-conditioned response generation function.


\subsection{Expertise inference via semantic-behavioral cues}
\label{sec:keystroke}
We estimate expertise levels using in-context learning, taking into account both the semantic context of the user query and behavioral signals such as keystroke dynamics during query composition. Our method includes two key components: (i) a feature extraction pipeline that captures both the user's semantic query context and behavioral context, converting key-level keystroke events into interpretable temporal statistics at the word level; and (ii) a prompt-based inference mechanism that uses feature representations, along with few-shot examples, to classify the user's query domain expertise.

\begin{figure*}[t]
    \captionsetup{labelfont={color=revcolor}, textfont={color=revcolor}}
    \centering
    \includegraphics[width=\linewidth]{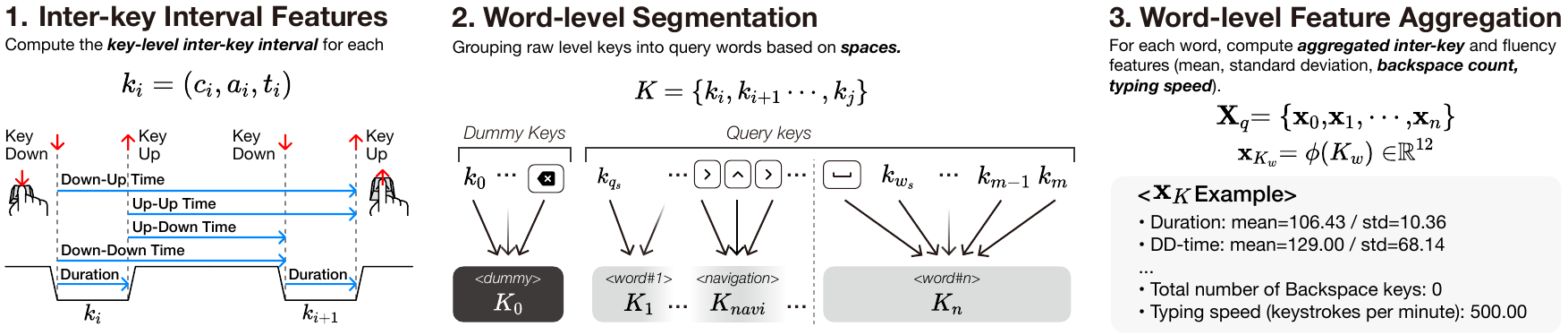}
    \caption{Overview of the keystroke feature extraction pipeline. The process involves (i) computing inter-key interval features at the key level, (ii) segmenting raw keystrokes into word-level units based on space and word-start characters, and (iii) aggregating temporal and fluency features for each word.}
    \label{fig:keyfeature}
\end{figure*}

\paragraph{Feature extraction.}
We use both semantic and behavioral signals as key features, motivated by the intuition that both signals provide complementary evidence of a user's domain expertise. On the semantic side, the content and phrasing of a query reflect the user's conceptual understanding and familiarity with domain-specific terminology~\cite{10.1145/3411764.3445648}.
On the behavioral side, keystroke dynamics capture how confidently and fluently a user expresses their query~\cite{mao2018does}. 
We hypothesize that the synergetic integration of both signals enables more accurate expertise inference.
To capture semantic indicators of expertise, we include the raw query text \( q \) as part of the inference input. Rather than explicitly extracting linguistic features, we allow the LLM to interpret the semantic content of the query.

For capturing behavioral cues, illustrated in Figure~\ref{fig:keyfeature}, we encode keystroke dynamics using seven temporal and fluency features, adapted from prior work on user state and input behavior modeling~\cite{epp2011identifying, kolakowska2015recognizing, nahin2014identifying}. 
Specifically, let \( K_q = \{k_0, k_1, \dots, k_m\} \) denote the sequence of keystroke events for a given query, where each event \( k_i = (c_i, a_i, t_i) \) consists of key character \( c_i \), action type \( a_i \in \{\texttt{down}, \texttt{up}\} \), and timestamp \( t_i \). From this sequence, we compute five temporal features:
\textit{Duration}, which captures the hold time between key press and release for individual keys; and inter-key interval features, \textit{Down-Down}, \textit{Up-Down}, \textit{Up-Up}, and \textit{Down-Up}, which measure temporal intervals between consecutive key transitions. Additionally, we incorporate two global indicators of fluency: the total number of \textit{Backspace} keys, and \textit{Typing Speed}, defined as the number of keystrokes per minute.

While key-level features capture fine-grained timing, segmenting them at the word level yields stable behavioral units and aligns typing patterns with semantic content for expertise inference. To achieve this, we segment the raw keystroke stream into words using space and punctuation boundaries, identifying word starts by detecting characters preceded by a space key. 
For each word \(w\), we compute a word-level feature vector from the corresponding keystroke events \(K_w\). 
Specifically, we calculate the mean and standard deviation for one duration feature and four inter-key interval features, resulting in \(2 \times 5 = 10\) dimensions, and include two additional fluency features. 
This yields a 12-dimensional representation defined as:
\[
\mathbf{x}_{K_w} = \phi(K_w) \in \mathbb{R}^{12},
\]
where \(\phi\) aggregates the above statistics into the 12-dimensional feature space.

To account for typing behaviors that occur before the user begins inputting the first character of a query, we introduce a \textit{dummy group} represented by \(\mathbf{x}_0 = \phi(K_0)\), 
where \(K_0\) includes all keystrokes that occur prior to the first typed character and where the immediately preceding keystroke is a backspace. While the dummy group is not included in the natural language query passed to the LLM, it is treated identically during feature extraction to preserve early behavioral signals. \rev{In addition, we define a \textit{navigation group} for keystroke events related
to cursor movement (e.g., arrow keys) during query formulation, which captures revision behavior
and is retained as part of the behavioral signal.

Therefore, given a query \(q\) consisting of a dummy group, word-level groups, and navigation groups, the complete behavioral representation is:
\[
\mathbf{X}_q = \{ \mathbf{x}_i \}_{i=1}^{|\mathcal{G}_q| },
\]

where each \(\mathbf{x}_i = \phi(K_i)\) corresponds to a behavioral group (dummy, word-level, or navigation) extracted from the keystroke sequence, and \(|\mathcal{G}_q|\) denotes the total number of groups for query \(q\).

} 
Finally, the input pair used for the expertise inference prompt is defined as:
\[
I_q = \{q, \mathbf{X}_q\}.
\]
Additional details and examples are in Appendix~\ref{appendix:feature_extraction}.


\paragraph{Expertise inference prompt.}
We frame the expertise inference task as an in-context classification problem~\cite{min2022rethinking, zhang2022active}, allowing the LLM to infer user expertise directly from structured semantic-behavioral features. To support this, we design a system prompt composed of four key components: (i) a role and task definition, including an explanation of the five-level expertise taxonomy; (ii) natural language descriptions of temporal features extracted from keystroke logs; (iii) explicit multi-step \rev{classification procedure} that guide the model's decision-making process; and (iv) a set of few-shot examples comprising input-output pairs.



To enable expertise inference, we construct a few-shot example. Let \( Q = \{q_0, q_1, \dots, q_M\} \) denote the full set of queries. We then select a subset of \( N \) queries \( \{q_0, q_1, \dots, q_N\} \subset Q\), where each query is represented as an input pair $I_{q_i} = \{q_i, \mathbf{X}_{q_i}\}$ and annotated with a self‑reported expertise label $e_{q_i} \in \{\text{Novice}, \dots, \text{Expert}\}$. 
Each example in the prompt is formatted as a structured input-output pair:
\[
\mathcal{D}_i = (I_{q_i},\ e_{q_i}).
\]
The final in-context support set is:
\[
\mathcal{S} = \{ \mathcal{D}_1, \dots, \mathcal{D}_N \}.
\]
where \( N = 10 \) in our default configuration, a design choice made to balance context richness with prompt length in the absence of strong prior constraints~\cite{chen2023many, wu2023scattershot}.

Finally, we append the user prompt---comprising the current query and keystroke features, enabling inference of the user's expertise. 


\subsection{Expertise-conditioned response generation}
\label{sec:response}

To personalize LLM responses, we design an \textit{expertise injection prompt} that encodes the user's query domain expertise within the prompt input.
This framework guides the model to align the level of detail, terminology, and conceptual complexity of its responses with the user's expertise, enabling more reliable personalization of tone and depth.


\paragraph{Expertise injection prompt.}
We construct a system prompt with two main components: 
(i) structured metadata specifying the user's expertise level on a five-point taxonomy---Novice, Beginner, Intermediate, Proficient, and Expert ~\cite{palta2025speakingrightlanguageimpact}; and (ii) an expertise-level behavioral profile with generation guidance for calibrating response complexity, terminology, and abstraction. Additionally, the user query is included as a user prompt.



\section{Data Collection}
\label{sec:datacollection}
To construct a query-level user behavior dataset for expertise-conditioned response generation and satisfaction evaluation, we conducted a user study that recorded keystroke dynamics, query text, self-reported expertise levels, and response satisfaction.
More details are provided in Appendix~\ref{appendix:userstudy}.

\paragraph{System.}
We developed a custom web-based chatbot that supports querying an LLM, collecting query text, satisfaction ratings, and fine-grained keystroke dynamics. Responses were generated using \texttt{chatgpt-4o-latest}~\cite{openai2024chatgpt4o}.

\paragraph{Study design and procedure.}
Participants completed three domain-specific tasks in chemistry, computer science, and business, chosen to span varying levels of domain familiarity and conceptual complexity. For each domain, a unique task was designed, and participants used our chatbot to investigate the task. For each query, they reported perceived expertise (used for response generation) and rated response satisfaction on a 5-point Likert scale. After each task, participants completed a post-task survey assessing self-reported measures, followed by a short multiple-choice quiz.



\paragraph{Participants.}
\rev{We recruited 40 participants (20 males, 20 females, mean age of 23.10 (SD 1.61)), targeting students with different academic backgrounds. To ensure sufficient expertise, participants were selected from each domain: 13 in chemistry, 14 in computer science, and 13 in business.}


\paragraph{Dataset summary.}
For each query, we recorded the text by the user, the self‑reported expertise level, and two LLM responses (baseline and expertise-conditioned) along with corresponding satisfaction ratings, and raw keystroke data logged as sequential key presses and releases with timestamps. \rev{In total, the study yielded 1270 queries spanning the three domains, forming a dataset of textual, behavioral, and subjective user feedback. The dataset comprised 651 Novice, 245 Beginner, 209 Intermediate, 120 Proficient, and 45 Expert queries.}

\renewcommand{\arraystretch}{0.95}
\begin{table*}[t]
\color{revcolor}
\captionsetup{labelfont={color=revcolor}, textfont={color=revcolor}}
\centering
\caption{
Comparison of expertise inference performance under within-user evaluation with 10-shot in-context examples. Columns indicate the adaptation type (dynamic at the session level or static), the semantic and behavioral features used, and MAE and MSE, where lower values indicate better performance.
}
\label{tab:expertise-results}
\resizebox{0.85\linewidth}{!}{%
\begin{tabular}{lccccl}
\toprule
\textbf{Method} & \textbf{Adaptation} & \textbf{Semantic} & \textbf{Behavioral} & \textbf{MAE $\downarrow$} & \textbf{MSE $\downarrow$}  \\
\midrule
Random Guess & Static  & \xmark & \xmark & 1.600 & 4.000 \\
Persona-based~\cite{hu-collier-2024-quantifying} & Static  & \cmark & \xmark & 1.251 & 2.341 \\
Session-level~\cite{palta2025speakingrightlanguageimpact} & Static & \cmark & \xmark & 1.307 & 2.406 \\
IDL~\cite{cheng2024dialogues} & Dynamic  & \cmark & \xmark & 1.162 & 1.885 \\
AI Persona~\cite{wang2024ai} & Dynamic & \cmark & \xmark & 1.353 & 2.633 \\
Semantic-only & Dynamic  & \cmark & \xmark & 0.488 & 0.858 \\
Behavior-only & Dynamic  & \xmark & \cmark & 0.790 & 1.600 \\
\textbf{\system{}} (Semantic+Behavior) & Dynamic  & \cmark & \cmark & \textbf{0.398} & \textbf{0.698} \\
\bottomrule
\end{tabular}
}    
\label{tab:mainresults}
\end{table*}

\section{Expertise Level Inference}
We evaluate our expertise inference method, which estimates user expertise per query from semantic content and keystroke features without long-term interaction history or explicit user profiles.

\subsection{Setup}

\paragraph{Baselines.}
\label{sec:baselines}
As a theoretical lower bound, we consider a \textbf{Random Guess} baseline that is a random predictor uniformly sampling labels from the five-point expertise scale. We then compare our semantic-behavioral inference method with prior approaches. We include a \textbf{Persona-based Classifier} baseline based on Hu et al.~\cite{hu-collier-2024-quantifying}, which uses zero-shot prompting with static user attributes (e.g., demographics and attitudes) to predict subjective labels, serving as a reference for profile-based personalization without dynamic or behavior-aware adaptation.
Second, we implement a \textbf{Session-level Classifier} inspired by prior work~\cite{palta2025speakingrightlanguageimpact}, which aggregates users' past queries and model responses within a session to predict expertise, capturing long-term interaction context without per-query or behavior-aware adaptation.
\rev{Lastly, we implement two \textbf{Dynamic Adaptation} baselines that evolve with changes in the user-system dialogue. In-Dialogue Learning (IDL)~\cite{cheng2024dialogues} extracts persona-related sentences from user dialogue for fine-tuning and DPO-based adaptation, while AI Persona~\cite{wang2024ai} dynamically updates persona attributes during conversation to enable real-time profile adjustment. }More details are provided in Appendix~\ref{appendix:baselines}.

\paragraph{Metrics.}
We evaluated performance based on the mean absolute error (MAE) and mean squared error (MSE) between the predicted and self-reported expertise levels. Expertise labels were mapped to a five-point scale (Novice=1, Expert=5).


\subsection{Results}
\label{sec:results}

\begin{figure}[t]
    \captionsetup{labelfont={color=revcolor}, textfont={color=revcolor}}
    \centering
        \begin{subfigure}{0.48\linewidth}
            \includegraphics[width=\linewidth]{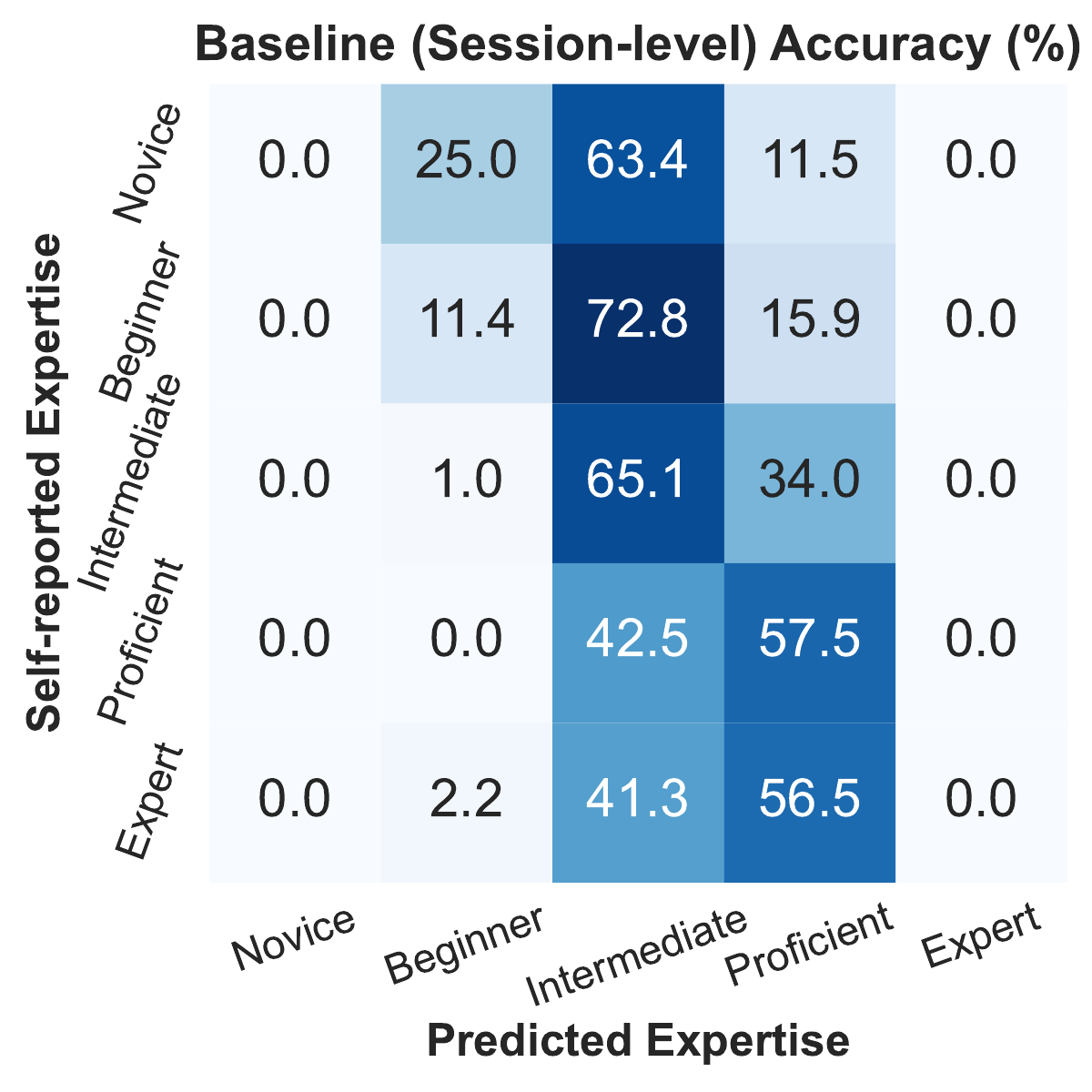}
        \end{subfigure}
        \begin{subfigure}{0.48\linewidth}
            \includegraphics[width=\linewidth]{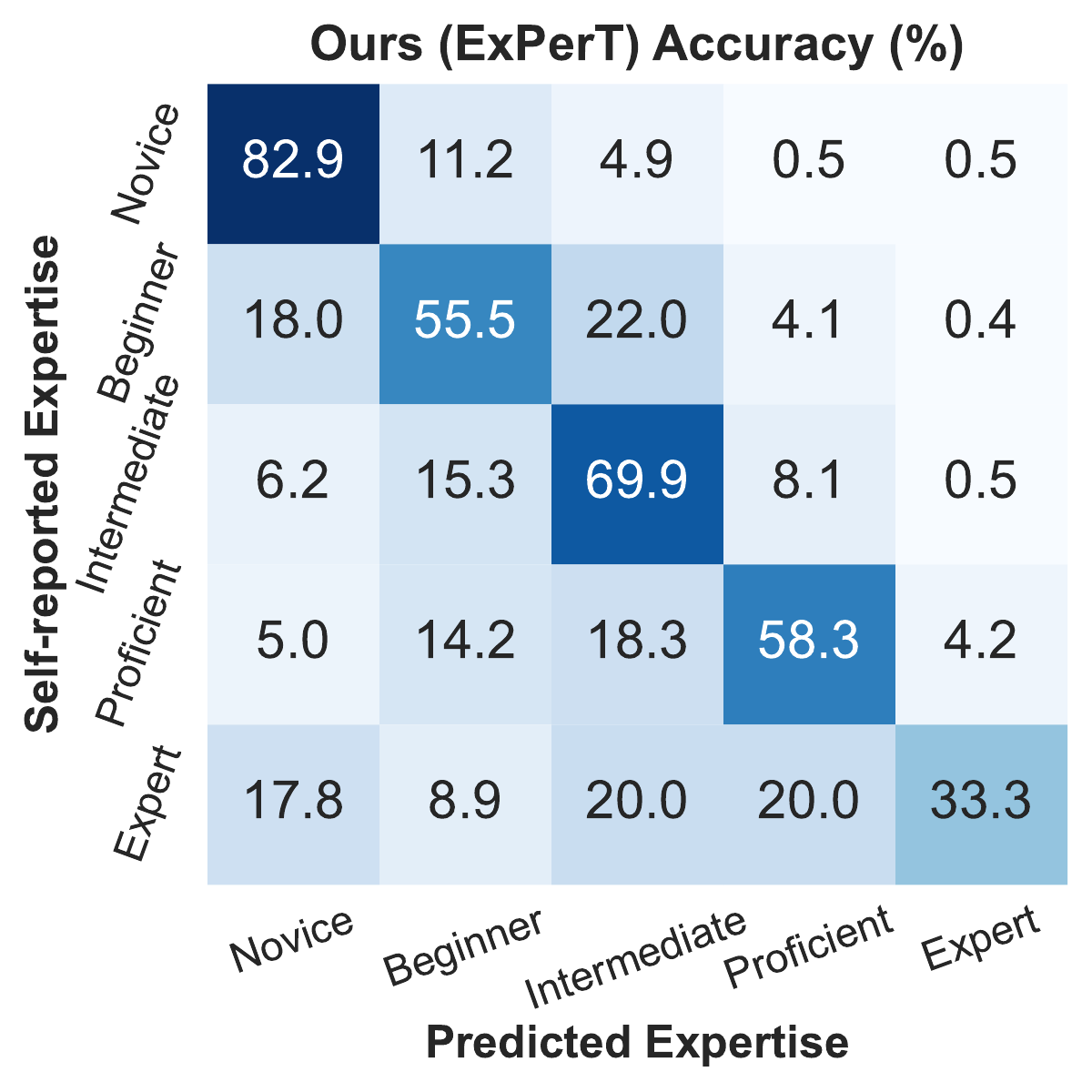}
        \end{subfigure}
    \caption{
    Confusion matrices comparing predicted and self-reported expertise levels; left: the session-level baseline~\cite{palta2025speakingrightlanguageimpact}, right: our method (\system{}).
    }
    \label{fig:confusion_matrix}
\end{figure}

We evaluated keystroke-based expertise inference by comparing model predictions to users' self-reported expertise levels. Inference was conducted in the within-user setting, using 10-shot examples drawn from the same user, unless specified, with GPT-4.1~\cite{openai2024gpt41}.
As shown in Table~\ref{tab:expertise-results}, \system{} achieved the best overall performance with a mean absolute error (MAE) of 0.398 and a mean squared error (MSE) of 0.698.
\rev{\system{} outperformed the strongest baseline, IDL~\cite{cheng2024dialogues}, 
reducing the MAE by 65.7\% (from 1.162 to 0.398).} 
As shown in Figure~\ref{fig:confusion_matrix}, we further compare \system{} with the session-level baseline, which explicitly predicts expertise. While the session-level baseline predictions were heavily concentrated on the intermediate expertise level, our method produced a more balanced distribution that better captured the full range of user expertise.


\begin{table*}[t]
    \color{revcolor}
    \captionsetup{labelfont={color=revcolor}, textfont={color=revcolor}}
    \centering
    \scriptsize
    \caption{Case study of semantic-behavioral complementarity: two queries with contrasting expertise (left: computer science, Expert; right: chemistry, Beginner).
    Semantic-only underestimates an expert's open-ended request, while Behavior-only overestimates a beginner due to fluent typing; \system{} combines both cues to match user expertise.
    }
    \renewcommand{\arraystretch}{1.1}
    
    \begin{tabularx}{0.98\textwidth}{
      p{1.8cm}
      >{\raggedright\arraybackslash}X
      >{\raggedright\arraybackslash}X
    }
        \toprule
         Query & Tell me everything that you know about priority scheduling and multilevel queue scheduling  & Can you explain in detail correlation with spectroscopic properties(e.g. IR and Raman activity)? \\
        \midrule
        Reported expertise & {\centering\textbf{Expert}} & {\centering\textbf{Beginner}} \\
        \midrule
        \textbf{ExPerT (ours)}
        & \textbf{Prediction:} Expert. \newline \textbf{Reason:} \ldots  The user uses precise technical terms (""priority scheduling,"" ""multilevel queue scheduling"") without hesitation or correction in those segments, as shown by stable keystroke metrics (low backspace, consistent timing, high typing speed). This matches the pattern in Example 7, where similar open-ended, technical queries and fluent typing are associated with Expert-level users. \ldots
        & \textbf{Prediction:} Beginner. \newline \textbf{Reason:} Baseline from prior queries (see Example 10) is Beginner, with consistent requests for careful explanation and explicit self-identification as a beginner. \ldots Typing behavior on technical segments is stable (no strong hesitation or correction), so no negative adjustment is warranted, but there is also no evidence for higher conceptual mastery or application. Final level remains Beginner. \\
        \midrule
        Semantic-only
        & \textbf{Prediction:} Novice. \newline \textbf{Reason:} \ldots The user requests ""everything"" about the topics, which is characteristic of someone with little or no familiarity seeking a general overview.
        & \textbf{Prediction:} Beginner. \newline \textbf{Reason:} \ldots The phrasing suggests curiosity and a need for foundational explanation, consistent with a beginner level. \\
        \midrule
        Behavior-only
        & \textbf{Prediction:} Proficient. \newline \textbf{Reason:} \ldots The user demonstrates fluent, confident typing with low error correction, and the timing metrics (mean and std) are within the range of proficient or expert users in the examples \ldots
        & \textbf{Prediction:} Proficient. \newline \textbf{Reason:} \ldots  user demonstrates fluent, confident typing with low error correction, and the timing metrics (mean and std) are stable and similar to those seen in proficient or expert examples. \ldots \\
        \bottomrule
    \end{tabularx}
    \label{tab:case_study}
\end{table*}
\begin{figure*}[t]
\captionsetup{labelfont={color=revcolor}, textfont={color=revcolor}}
    \centering
    \includegraphics[width=\linewidth]{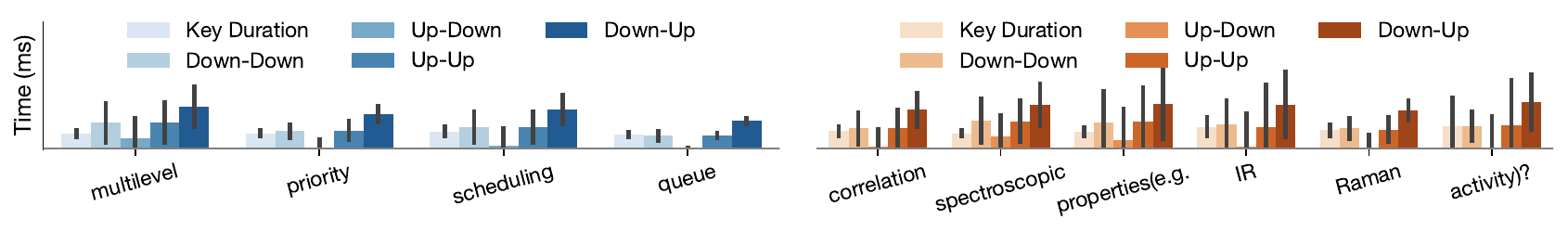}
    \caption{Inter-key interval patterns on domain-specific terms across expertise levels. Left: Expert; right: Beginner. Experts show both smaller intervals and lower variability (smaller STD) than beginners in domain-specific terms.}
    \label{fig:indepth_fig}
\end{figure*}
\begin{figure}[t]
\captionsetup{labelfont={color=revcolor}, textfont={color=revcolor}}
    \centering
        \begin{subfigure}{\linewidth}
            \includegraphics[width=\linewidth]{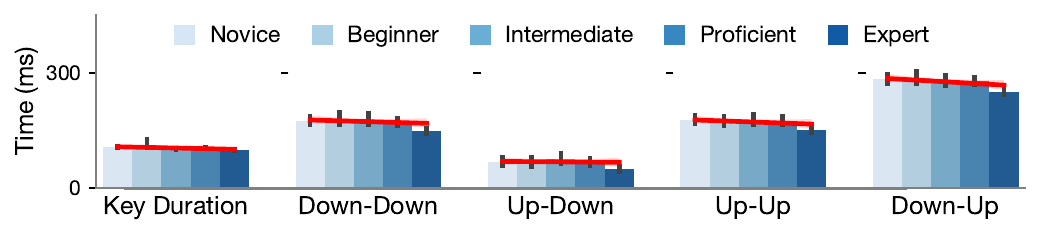}
        \end{subfigure}
        \begin{subfigure}{\linewidth}
            \includegraphics[width=\linewidth]{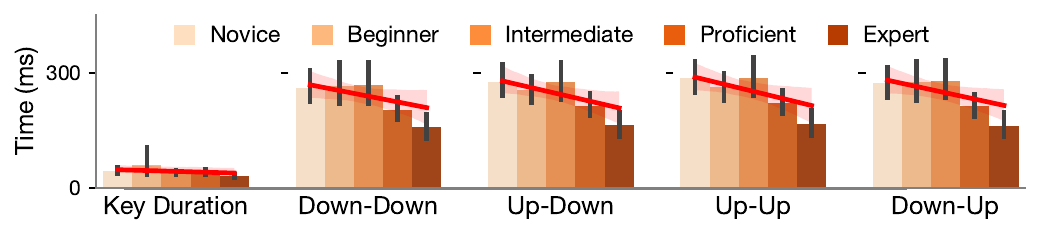}
        \end{subfigure}
    \caption{Top: inter-key intervals for domain-specific terms across different expertise levels; bottom: word-level standard deviations of inter-key intervals. Higher expertise shows smaller intervals and lower variability.}
    \label{fig:rawkey_feature}
\end{figure}
\rev{\paragraph{Complementary advantages of semantic and behavioral features.}

\label{sec:synergetic}
Beyond aggregate performance improvements, a fine-grained analysis is provided to clarify how semantic and behavioral cues are complementary for expertise inference.
Semantic cues indicate what technical terms appear and how they are used in context (e.g., definition-seeking vs.\ conceptual use), while keystroke behavioral cues capture how fluently users produce those same terminologies during composition.

Figure~\ref{fig:rawkey_feature} shows systematic expertise-aligned trends across multiple inter-key intervals: as expertise increases, both mean intervals and word-level variability decrease on domain-specific terms. 
Figure~\ref{fig:indepth_fig} illustrates this effect for each example, showing that experts type domain-specific terms faster and more consistently, whereas beginners exhibit slower and more variable timing when composing them.
These patterns collectively show how timing-based fluency provides an orthogonal signal that complements the semantic interpretation of term usage, thereby reducing ambiguity in expertise inference. Statistical analyses supporting these observations are provided in Appendix~\ref{appendix:indepthkey_1}.

Crucially, \system{} infers expertise by jointly leveraging semantic and behavioral cues, which provide complementary evidence. As shown in Table~\ref{tab:case_study}, left example, stable keystroke behaviors on segments (e.g., low backspace rate and consistent inter-key timing) support an expert-level interpretation for the domain-specific terms ("priority scheduling", "multilevel queue scheduling"), leading \system{} to classify as \textit{Expert}. In contrast, semantic-only classifies the same query as \textit{Novice} by treating the broad phrasing ("tell me everything") as a low-expertise overview request.
In the right example, a query about spectroscopic properties is retained as \textit{Beginner} despite fluent typing, because the domain-specific terms (e.g., "IR" and "Raman activity") are used only to request an explanation rather than to apply or integrate concepts. In contrast, behavior-only inference labels it as \textit{Proficient} due to low error correction and stable timing.

These cases demonstrate complementarity: combining semantic and behavioral cues corrects semantic-only misinterpretations of open-ended phrasing and behavior-only overestimation driven by fluent typing. (See Appendix~\ref{appendix:expertise_inference_qual} for full cases.)
}

\begin{figure}[t]
    \captionsetup{labelfont={color=revcolor}, textfont={color=revcolor}}
    \centering
    \includegraphics[width=\linewidth]{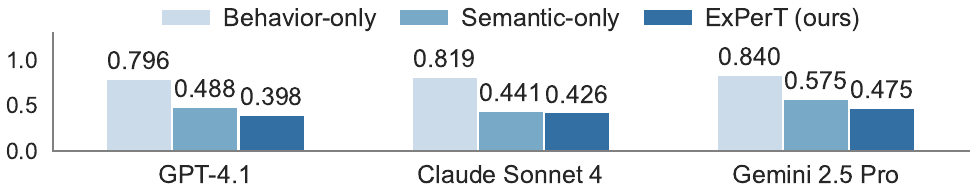}
    \caption{Comparison of MAE in expertise inference among GPT-4.1, Claude Sonnet 4, and Gemini 2.5 pro.}
    \label{fig:model_generalization}
\end{figure}

\begin{table}[t]
\color{revcolor}
\captionsetup{labelfont={color=revcolor}, textfont={color=revcolor}}
\centering
\caption{Ablation results on the effect of the dummy group in ExPerT (within-user, 10-shot examples).}
\label{tab:dummy}
\resizebox{0.9\columnwidth}{!}{%
\begin{tabular}{lcc}
\toprule
\textbf{Keystroke Representation}  & \textbf{MAE $\downarrow$} & \textbf{MSE $\downarrow$} \\
\midrule
ExPerT w/o dummy group & \rev{0.403} & \rev{0.715}\\
ExPerT with dummy group & \rev{\textbf{0.398}} &  \rev{\textbf{0.698}}\\
\bottomrule
\end{tabular}
}
\end{table}

\begin{table}[t]
\centering
\caption{Ablation results on different keystroke representation methods (within-user, 10-shot examples).}
\label{tab:ablation2}
\resizebox{0.9\columnwidth}{!}{%
\begin{tabular}{lcc}
\toprule
\textbf{Keystroke Representation}  & \textbf{MAE $\downarrow$} & \textbf{MSE $\downarrow$} \\
\midrule
Key-level Raw Feature & \rev{0.483} & \rev{0.887}\\
Query-level Aggregation & \rev{0.410} & \rev{0.735}\\
Word-level Aggregation & \rev{\textbf{0.398}} &  \rev{\textbf{0.698}}\\
\bottomrule
\end{tabular}
}
\end{table}

\rev{
\paragraph{Model generalizability.}
\label{sec:model generalizability}

To assess generalizability beyond a single model, we evaluate \system{}'s performance across multiple LLMs, including GPT-4.1~\cite{openai2024gpt41}, Claude Sonnet 4~\cite{anthropic2025claude4}, and Gemini 2.5 Pro~\cite{comanici2025gemini25pushingfrontier}. As shown in Figure~\ref{fig:model_generalization}, \system{} consistently outperforms semantic-only and behavioral-only across all models, indicating model-agnostic effectiveness. 
}


\rev{\paragraph{Dummy group.}
\label{sec:dummygroup}
We conduct an ablation study to examine the effect of the dummy group. As shown in Table~\ref{tab:dummy}, removing the dummy group degrades performance, with MAE increasing by 1.3\% (from 0.398 to 0.403). It suggests that the dummy group provides a stabilizing reference during inference.
}

\paragraph{Keystroke representation methods.}
To evaluate how keystroke representations affect expertise inference in \system{}, we compared three approaches: (i) raw key-level features without aggregation, (ii) query-level features aggregated over the entire query, and (iii) word-level features aggregated per word (\system{}). As shown in Table~\ref{tab:ablation2}, word-level aggregation achieves the best performance, suggesting that it better preserves behavioral patterns than either unstructured raw keystrokes or coarse query-level statistics (see Appendix~\ref{appendix:feature_extraction}).


\begin{figure}[t]
\captionsetup{labelfont={color=revcolor}, textfont={color=revcolor}}
    \centering
    \includegraphics[width=\linewidth]{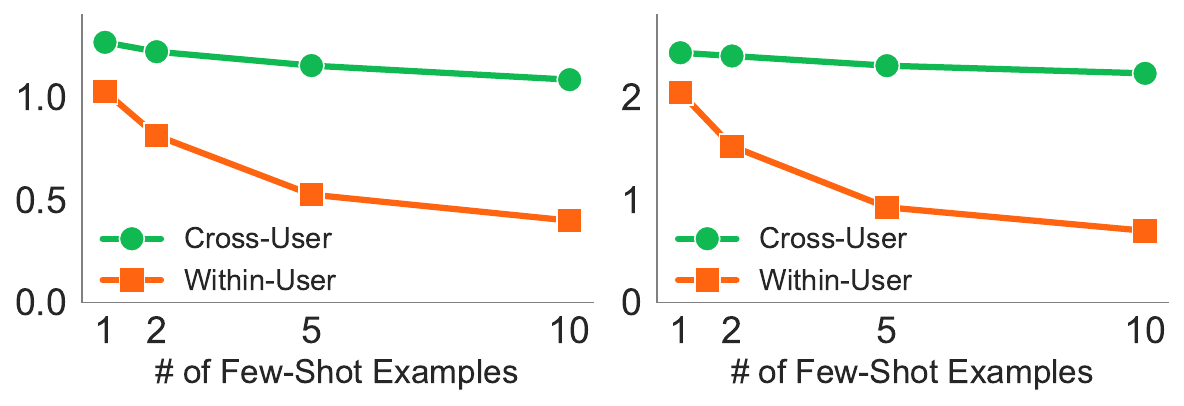}
    \caption{Left: MAE and right: MSE of expertise inference according to the number of few-shot examples.}
    \label{fig:ablation-fewshot}
\end{figure}

\paragraph{Few-shot examples.}
To assess few-shot configuration effects, we vary the number of examples (1, 2, 5, and 10) and the sampling source (within-user vs. cross-user). ExPerT supports both settings: within-user personalization when prior interactions from the same user are available, and cross-user inference for cold-start settings where such history is unavailable. As shown in Figure~\ref{fig:ablation-fewshot}, performance improves with more examples, with the within-user setting achieving the best results at 10-shot, which suggests that even a modest number of examples is sufficient for accurate inference. 
In contrast, the cross‑user setting, where all few‑shot examples were drawn from different users, performed worse, likely due to misaligned behavioral patterns~\cite{gao2024noise}. This gap highlights the importance of user-specific cues, as typing dynamics tend to vary significantly across individuals~\cite{plank2016keystroke, keyr4, keyr5, keyr6}. \rev{Additional statistical analyses of user-dependency are reported in Appendix~\ref{appendix:indepthkey_2}.}



\begin{figure}[t]
\captionsetup{labelfont={color=revcolor}, textfont={color=revcolor}}
    \centering
        \begin{subfigure}{\linewidth}
            \includegraphics[width=\linewidth]{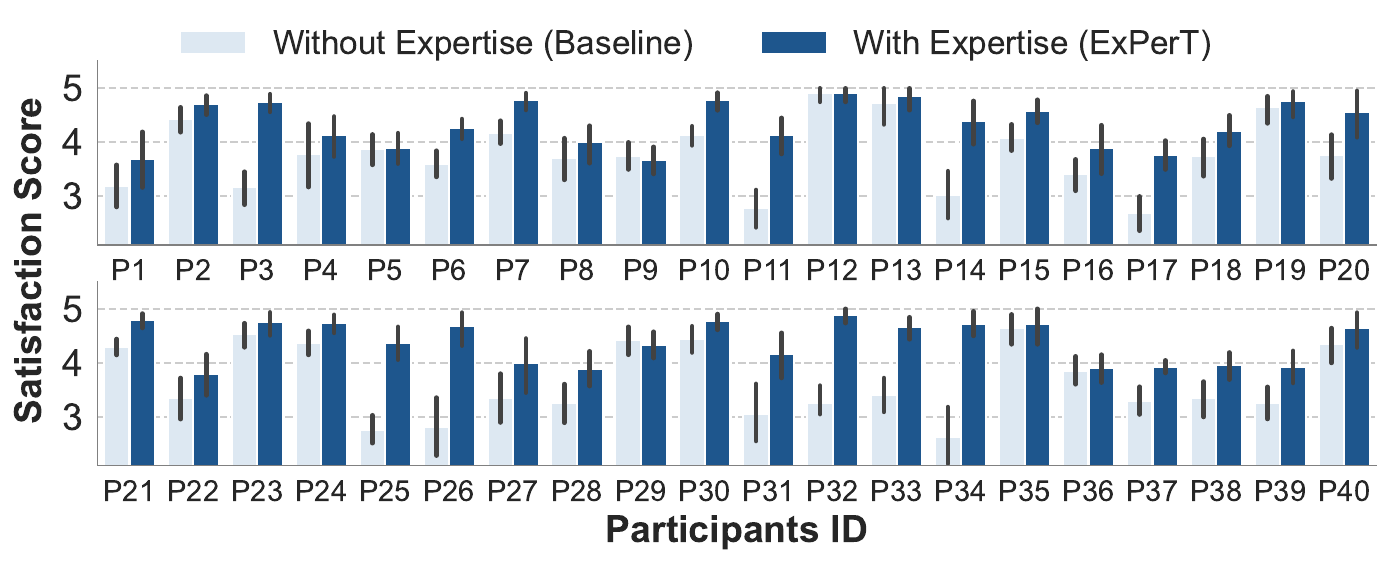}
        \end{subfigure}
        \begin{subfigure}{\linewidth}
            \includegraphics[width=\linewidth]{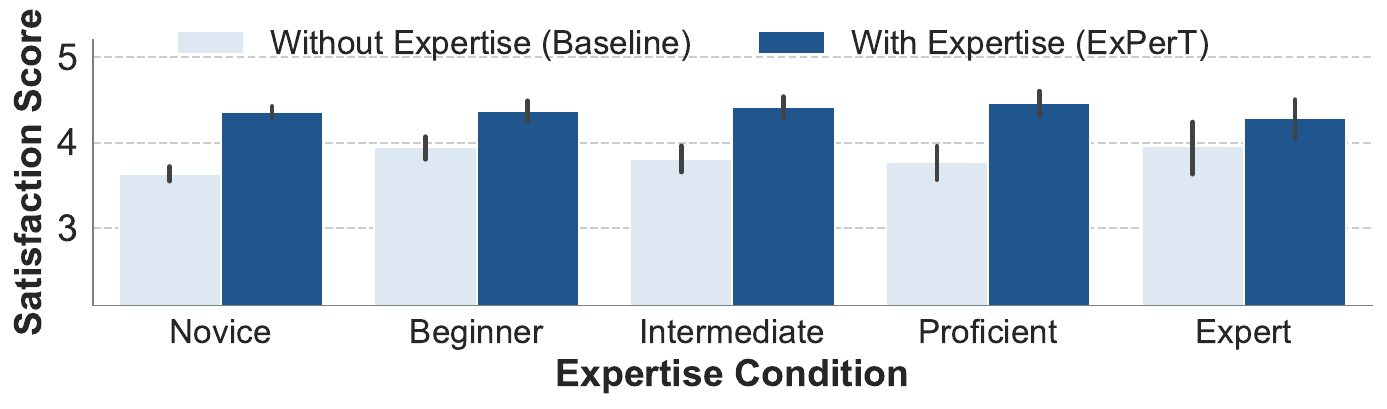}
        \end{subfigure}
    \caption{User satisfaction scores for baseline (without expertise) and expertise-conditioned responses (with predicted expertise). Top: grouped by participants; bottom: grouped by self-reported expertise level.}
    \label{fig:satf_expertlevel}
\end{figure}

\section{Expertise-Conditioned Response Generation}

\cm{In this section, we report two response-generation evaluations. Together, these evaluations assess whether conditioning responses on user expertise improves user satisfaction (Section~\ref{sec:satisf-study1}) and whether these benefits persist when expertise is inferred rather than self-reported (Section~\ref{sec:satisf-study2}).}

\subsection{Self-Reported Expertise Conditioned User Satisfaction}
\label{sec:satisf-study1}
We analyze the expertise-conditioned responses collected during the data-collection stage, stratified by participants' self-reported expertise. Because the data violated the normality assumption, we use the Wilcoxon signed-rank test for paired comparisons.


\paragraph{User satisfaction}
\label{sec:user-satisfaction}
\rev{
As illustrated in Figure 9, expertise-conditioned responses received significantly higher satisfaction ratings than baseline responses. The mean score increased from $3.71$ ($SD = 0.64$) to $4.36$ ($SD = 0.40$), corresponding to an average improvement of $0.65$ ($SD = 0.56$) ($p<0.001$, $W = 51824.0$, $Z=-26.94$, $r=0.76$).
Most participants rated \system{} higher than the baseline, with higher satisfaction across all self-reported expertise levels.
}

\paragraph{User expertise dynamics}
\label{sec:user-expertise-dynamics}
Participants' self-reported expertise varied substantially across and within domains, depending on subtopic and recent exposure, highlighting its context-dependent nature. Moreover, expertise assessments fluctuated within sessions, as participants revised their perceived expertise upward with growing familiarity or downward when encountering unfamiliar concepts or unexpected (p1, p11, p12, p13, p17). These findings underscore the dynamic nature of expertise and motivate adaptive LLMs that tailor responses to users' evolving cognitive states. (See Appendix~\ref{appendix:survey-results}.)

\cm{
\subsection{Follow-up Evaluation with Predicted Expertise}
\label{sec:satisf-study2}

We present a follow-up study with returning participants, who evaluated three responses to queries they had originally submitted: a response without expertise conditioning, one conditioned on self-reported expertise, and one conditioned on ~\system{}-predicted expertise (See Appendix ~\ref{sec:followup}).

\paragraph{Study and participants.}
For the follow-up study, we recruited 16 returning participants (5 males, 11 females, mean age of 22.63 (SD = 1.45)) from the original data collection stage. Participants were selected from each domain: 7 in chemistry, 7 in computer science, and 2 in business. All responses were generated using \texttt{gpt-5.3-chat-latest}~\cite{openai2026chatgpt53}. For each query, participants were shown three responses and independently rated their satisfaction with each response on a 5-point Likert scale. 

\paragraph{User satisfaction}

\begin{table}[t]
\centering
\setlength{\tabcolsep}{4pt}
\resizebox{\linewidth}{!}{%
\begin{tabular}{cccccc}
\toprule
 & Novice  & Beginner  & Intermediate  & Proficient  & Expert  \\
 & (n=291) & (n=72) & (n=90) & (n=44) & (n=25) \\ 
\midrule
Without & 3.60 & 3.49 & 3.66 & 3.57 & 3.52 \\
Reported & \textbf{4.08} & \textbf{4.15} & 3.67 & 3.89 & \textbf{4.28} \\
~\system{} & 4.02 & 3.96 & \textbf{3.83} & \textbf{3.91} & 4.12 \\
\bottomrule
\end{tabular}%
}
\caption{Mean satisfaction ratings across response conditions, grouped by participants' self-reported expertise level. Higher scores indicate greater satisfaction.}
\label{tab:mean_response_by_expertise}
\end{table}


In this study, we collected 522 satisfaction queries.
As illustrated in Table~\ref{tab:mean_response_by_expertise}, expertise-informed responses increased user satisfaction: self-reported expertise received the highest ratings ($M = 4.010, SD = 0.947$), predicted expertise showed similar ratings ($M = 3.975, SD = 0.982$), and no-expertise received the lowest ratings ($M = 3.586, SD = 0.994$). A repeated-measures ANOVA showed a significant effect of response type ($F(2, 30) = 28.39, p < .001, \eta^2 = .208$). Holm-corrected post hoc comparisons showed that both the self-reported expertise condition ($t(15) = 6.09, p < .001$) and the predicted expertise condition ($t(15) = 5.79, p < .001$) were rated significantly higher than the no-expertise condition, with no significant difference between the self-reported and predicted expertise conditions ($t(15) = 1.22, p = .241$). These results indicate that \system{}-predicted expertise achieved user satisfaction comparable to self-reported expertise.


}
\section{Conclusion}
We introduce \system{}, a query-wise personalization framework that adapts LLM responses to users' expertise via semantic and behavioral cues. Our inference method achieves an MAE of 0.398, surpassing both session-level and persona-based baselines.  and improves user satisfaction. 

\cm{
\section*{Acknowledgments}
This work was supported by Institute of Information \& communications Technology Planning \& Evaluation(IITP) grant funded by the Korea government(MSIT)(No.RS-2020-II201336, Artificial Intelligence graduate school support(UNIST)), (RS-2025-25442824, AI Star Fellowship Program(Ulsan National Institute of Science and Technology)), and under the Leading Generative AI Human Resources Development(IITP-2026-RS-2024-00360227). This work was also supported by the National Research Foundation of Korea (NRF) grant funded by the Korea government (MSIT) (RS-2025-00553241). In addition, this research was supported by the "Advanced GPU Utilization Support Program" funded by the Government of the Republic of Korea (Ministry of Science and ICT).

We also thank the UAI Lab members, especially Changmin, Jaemin, and Jiseung, for their feedback and participation in the pilot study.

\section*{Limitation}
\label{sec:limitation}
While our in-context learning-based approach enables fine-grained personalization, it has several limitations. First, the method remains sensitive to prompt design and few-shot exemplar composition. In addition, because inference benefits from within-user demonstrations, its robustness may decrease in cross-user or cold-start settings where user-specific interaction history is unavailable.

Second, keystroke dynamics are informative but inherently noisy behavioral signals. Features such as typing speed, backspace rate, and temporal variation may reflect confounding factors including fatigue, keyboard familiarity, or device differences rather than expertise alone~\cite{kolakowska2015recognizing, epp2011identifying}. This may reduce robustness across heterogeneous users, devices, and usage contexts. In addition, behavioral signals such as keystroke dynamics raise privacy, fairness, and accessibility concerns, including residual identifiability risk and potential bias related to motor impairments, device familiarity, or input conditions. Future work should therefore investigate privacy-preserving deployment strategies such as on-device personalization, secure communication, and minimizing cross-user sharing of behavioral data.

Finally, our study was conducted with a relatively small and homogeneous participant pool, primarily university students. As a result, the generalizability of ~\system{} to broader populations and interaction settings remains to be validated. Also, our labels rely on self-reported expertise, which serves as a practical proxy for users' perceived query-wise expertise but may not always align with objective proficiency or task performance. 
}

\bibliography{custom}
\appendix
\newpage


\section{Related Work}

\paragraph{LLMs for personalized response generation.}
Most existing LLM personalization studies rely on static approaches, such as persona-conditioned prompting~\cite{hu-collier-2024-quantifying}, retrieval-augmented generation (RAG)~\cite{10.1145/3616855.3635845, sun-etal-2025-persona}, or reinforcement learning from human feedback (RLHF) based fine-tuning~\cite{10.5555/3737916.3739580, chen2025padpersonalizedalignmentllms}, which assume persistent profiles and fixed preferences. Persona and RAG-based methods depend on long-term user data, yet often face context overflow or cold-start issues~\cite{10.1145/3616855.3635845, sun-etal-2025-persona}, while RLHF-based approaches tend to optimize for average user preferences, missing individual variation~\cite{10.5555/3737916.3739580, chen2025padpersonalizedalignmentllms}. In contrast, our approach enables query-wise personalized responses to the target user's expertise without relying on static profiles or retraining.

\paragraph{Semantic and behavioral features for user interaction.}
Recent work tried to personalize LLMs using either semantic features, such as few-shot semantic prompts~\cite{rodriguez2024intentgptfewshotintentdiscovery} and reasoning over emotion and intent in dialogue~\cite{wang2023cue}, or behavioral features, including mobile sensing for cognitive context~\cite{10.1145/3675094.3677545} and keystroke logging to capture writing behaviors~\cite{wang2025scholawritedatasetendtoendscholarly}. However, semantic methods rarely address domain expertise, and behavioral approaches often require specialized hardware or focus on aggregate patterns. Inspired by these studies, our approach infers and adapts to users' expertise levels via the synergetic use of semantic and behavioral signals, achieving accurate and personalized responses.

\paragraph{Impact of expertise on LLM interactions.}
The expertise reversal effect~\cite{Kalyuga_2021} shows that instructional strategies effective for novices can hinder experts, who prefer succinct, information-dense content~\cite{10.1145/3544549.3585900}. For instance, novices benefit from contextual explanations and step-by-step support, while experts prefer concise text and quick access to definitions~\cite{10.1145/3411764.3445648, 10.1145/3544549.3585900}. Misalignment between system output and user expertise significantly reduces user satisfaction and task success~\cite{palta2025speakingrightlanguageimpact}. While aligning LLMs with user proficiency improves engagement~\cite{xu2025expertpromptinginstructinglargelanguage, palta2025speakingrightlanguageimpact}, prior work relies on explicit expertise cues~\cite{hu-collier-2024-quantifying} or post-hoc classification~\cite{palta2025speakingrightlanguageimpact,magister2024way},
rather than instantaneous expertise-level inference. Our approach addresses this gap by dynamically adapting responses to each query.

\section{Expertise Inference}
\label{appendix:expertise_inference}
\subsection{Implementation details}
All inference experiments used \texttt{gpt-4.1}~\cite{openai2024gpt41} with \texttt{temperature}=0.0, \texttt{top\_p}=1.0, and a maximum output length of \texttt{512} tokens to prevent response truncation. In our experiments, few‑shot samples were randomly selected using a fixed random seed of 0.

\subsection{Prompt structure for expertise inference}
\label{appendix:expertise_inference_prompt}
We provide the core prompt structure used for expertise inference. This prompt guides the model to classify user expertise based on semantic and behavioral cues, including query text and keystroke-derived features. It specifies the model's role, expertise level taxonomy, reasoning steps, and required output format. 
To support consistent reasoning, we prepend a task instruction that decomposes the classification process into a series of sub-steps inspired by Chain-of-Thought prompting~\cite{wei2022chain}.  \rev{These steps explicitly instruct the LLM to (i) infer a baseline expertise level from within-user support examples that reflects the user’s overall topic familiarity, (ii) analyze the intent and conceptual characteristics of the current query to compute a bounded adjustment, (iii) identify technical terms in the query, (iv) use keystroke features such as hesitation, corrections, and stability only to modulate this adjustment on technical-term segments, and (v) combine the baseline and the bounded adjustment to select the final category within the five-level expertise taxonomy. This structured reasoning encourages stable, user-consistent predictions grounded in behavioral evidence. The following boxes present the exact system and user prompt templates used in our experiments.}

\begin{tcolorbox}[breakable, title=System Prompt, colback=white, colframe=black]
\begin{CodeBlock}
<core_identity>
You are an AI classifier that assigns a user's topic expertise level using BOTH the query text and keystroke dynamics, with strict safeguards against overestimating expertise.
</core_identity>

<general_rules>
• NEVER use meta-phrases (e.g., "let me help you", "I can see that").
• ALWAYS be specific and grounded in observable evidence.
• Do NOT infer personal traits beyond what is required for classification.
• START IMMEDIATELY WITH THE ANSWER – ZERO INTRODUCTORY TEXT.
</general_rules>

<general_instructions>
Classify the user into exactly one category: Novice, Beginner, Intermediate, Proficient, Expert. Do not infer unobservable mental states.
</general_instructions>

<expertise_scale>
• Novice: A subject novice is a person who has little or no familiarity with the topic or domain of the conversation.
• Beginner: A subject beginner is someone who has little prior knowledge or experience in the topic or domain of the conversation.
• Intermediate: A subject intermediate is someone who has some basic knowledge or familiarity with the topic of the conversation, but not enough to be considered an expert or even proficient.
• Proficient: A subject proficient is someone who can apply relevant concepts and terminology from the conversation to different scenarios and problems.
• Expert: A subject expert is someone who has a deep and comprehensive understanding of the topic or field of the conversation and can use specialized terms and references to communicate their knowledge.
</expertise_scale>

<keystroke_feature_definitions>
You are provided with a description of keystroke dynamics time-based features. Use this information to understand typing behavior and how different time intervals are defined.

Time-based features are defined for two consecutive keystroke events:
• Duration: The time that the user holds a key in the down position
• Down-down time (DD-time): The time between the press of a key and the press of the subsequent key
• Up-down time (UD-time): The time between the release of a key and the press of the subsequent key
• Up-up time (UU-time): The time between the release of a key and the release of the subsequent key
• Down-up time (DU-time): The time between the press of a key and the release of the subsequent key

Aggregate features:
• Typing speed (keystrokes per minute)
• Backspace count: The total number of Backspace key presses, representing 
correction activity.

{GIVEN_DATA}
</keystroke_feature_definitions>

<lemma technical_term_guidance>
Treat "technical term" as "domain-specific term with a specialized meaning in the given topic."
A query may contain technical terms even if the user is a novice (e.g., copied from class notes). Therefore, technical-term presence alone is weak evidence.
</lemma technical_term_guidance>

<classification_procedure>
Step 0) Estimate a BASELINE expertise level from the example pool (PRIMARY).
  - Use within-user consistency: what level best matches the user's typical queries and behaviors?
  - This baseline represents the user's overall topic expertise, not the difficulty of one question.

Step 1) Analyze the CURRENT query intent and conceptual use to compute a small adjustment (DELTA).
  - Definition/meaning/confirmation requests usually imply DELTA = -1 (not a hard cap).
  - Concept-linking targeted questions imply DELTA = 0.
  - Application/constraints/edge cases can imply DELTA = +1.

Step 2) Identify technical terms in the current query (lexical metadata).

Step 3) Use keystroke features ONLY to adjust DELTA based on technical-term segments:
  - Strong hesitation/corrections on technical-term segments --> DELTA -= 1 (bounded).
  - Stable technical-term segments --> keep DELTA as is (do not upgrade solely from typing).

Step 4) Final level = clamp(BASELINE + DELTA, within [Novice..Expert]).
  - By default, DELTA must be in {-1, 0, +1}.
  - Do NOT drop more than 1 level below BASELINE unless there is overwhelming evidence of mismatch vs the user's own history.
</classification_procedure> 

<Examples>
{EXAMPLE_POOL_INSTRUCTION}
{EXAMPLES}
</Examples>

<output_format>
Output must be structured exactly in this format:

<classification>
  <level>One of: Novice, Beginner, Intermediate, Proficient, Expert</level>
  <reason>Brief explanation of the reasoning steps and observed indicators.</reason>
</classification>
</output_format>
\end{CodeBlock}
\end{tcolorbox}

\
\subsection{User prompt format}
\begin{tcolorbox}[breakable, title=User Prompt, colback=white, colframe=black]
\begin{CodeBlock}
Given the following user input, determine the user’s expertise level based on typing behavior.
Query: {USER_QUERY}
Keystroke Summary: {BEHAVIORAL_FEATURE_SUMMARY}
Expertise Level:
\end{CodeBlock}
\end{tcolorbox}

\subsection{Output examples}
\begin{tcolorbox}[breakable, title=Output Examples, colback=white, colframe=black]
\begin{CodeBlock}
[Novice]
<level>Novice</level>
<reason>Baseline expertise for this user in business/accounting topics is Novice, as seen in prior queries about consolidated financial statements and related concepts. The current query includes technical terms (""Elimination of investments in subsidiaries against equity"", ""Recognition of goodwill or bargain purchase gains"", ""Accounting treatment of non-controlling interests""), but this matches previous novice-level queries where such terms were present without evidence of conceptual manipulation. Typing behavior on technical-term segments (e.g., ""Elmination"", ""goodwill"", ""subsidiaries"", ""accounting"") shows frequent corrections (high backspace counts: 39 for ""accounting"", 12 for ""Elmination"", 5 for ""non-controlling"", 7 for ""treatment"") and slow, inconsistent typing speeds, indicating hesitation and lack of fluency with these terms. There is no evidence of application, synthesis, or concept-linking in the query intent—it's a task prompt, not a conceptual or applied question. Therefore, the expertise level remains Novice.</reason>

[Beginner]
<level>Beginner</level>
<reason>Baseline expertise is Beginner, as established by prior queries involving basic conceptual questions and requests for examples in molecular symmetry and spectroscopy. The current query asks for
representative molecular examples for IR or Raman activity, which is a typical beginner-level request focused on concrete instances rather than application or theory. Technical terms (""IR"", ""Raman Activation"", ""Molecular"") are present, but keystroke dynamics on these segments (e.g., ""IR"", ""Raman"") show stable and relatively fast typing with no significant hesitation or correction, supporting the baseline but not warranting an upgrade. No evidence of conceptual application or advanced linkage is present. Final level remains Beginner.</reason>

[Intermediate]
<level>Intermediate</level>
<reason>Baseline is Intermediate, as prior queries show consistent use of scheduling algorithm terms and conceptual linking (e.g., ""Also tell me about multilevel feedback queues"", ""If it's a preemptive strike type srtf...""). The current query lists multiple technical terms (fcfs, sjf, Priority scheduling, Round robin) and requests detailed explanation, which is a concept-linking/application type but not at the level of deep analysis or edge-case exploration. Typing on technical terms is stable, with only minor corrections and no strong hesitation, supporting the baseline. No evidence for proficiency or expert-level synthesis. Final level: Intermediate.</reason>

[Proficient]
<level>Proficient</level>
<reason>Baseline is Proficient or higher, as the user previously demonstrated expert-level queries in chemistry (Example 4) and proficient-level reasoning in finance (Example 3). The current query uses precise technical terms (""crystallographic symmetry"", ""spectroscopic symmetry"", ""대칭요소"") and requests a comparative listing, which is a concept-linking/application question (DELTA = 0 or +1). Typing of technical terms is stable with only minor corrections (e.g., 1 backspace for ""crystallographic"", 0 for ""spectroscopic""), and no strong hesitation is observed on these segments. The high backspace count is concentrated in <dummy> and not on technical terms, so it does not warrant a downgrade. Final level is Proficient, as the query structure and typing are consistent with advanced but not fully expert-level domain integration.</reason>

[Expert]
<level>Expert</level>
<reason>Baseline expertise is Expert, as established by prior queries involving advanced OS concepts, technical terminology, and fluent application-level questioning (e.g., ""How to prevent
starvation in multilevel queue scheduling?"", ""When implementing a queue, explain heap, list, and tree and tell me the pros and cons of each.""). The current query uses technical terms (""preemptive algorithm"", ""process yield"", ""implement"") and asks about implementation details, indicating application-level understanding (DELTA = +1). Typing on technical segments (""process"", ""yield"") is stable, with no hesitation or correction, supporting the baseline. Final level remains Expert.</reason>

\end{CodeBlock}
\end{tcolorbox}
    
\subsection{Keystroke representation methods}
The description of input features provided to the model varies depending on the chosen keystroke representation method. As described in Appendix~\ref{appendix:feature_extraction}, we prepare multiple instruction of semantic‑behavioral input, including word‑level, query‑level, and raw‑level keystroke summaries.

\begin{tcolorbox}[breakable, title=Word-Level Input Features , colback=white, colframe=black]
\begin{CodeBlock}
Given Data: In this context, the data was preprocessed at the word level. For each word, the five key features described above were computed in terms of their mean and standard deviation (std). In total, In addition to these, two supplementary metrics were also included: the total number of Backspace keys and the overall typing speed. As a result, a total of seven features were extracted, as follows:
Note that <dummy> refers to a special group representing keystrokes that occur before the first typed character of a query, and <navigation> refers to keystrokes associated with navigation actions (e.g., arrow keys) rather than actual character input.

• Duration (mean/std)
• DD-time (mean/std)
• UD-time (mean/std)
• UU-time (mean/std)
• DU-time (mean/std)
• Total number of Backspace keys
• Typing speed (keystrokes per minute)
\end{CodeBlock}
\end{tcolorbox}

\begin{tcolorbox}[breakable, title=Query-Level Input Features , colback=white, colframe=black]
\begin{CodeBlock}
Given Data: In this context, the data was preprocessed at the query level. As a result, a total of seven features were extracted, as follows:

• Duration (mean/std)
• DD-time (mean/std)
• UD-time (mean/std)
• UU-time (mean/std)
• DU-time (mean/std)
• Total number of Backspace keys
• Typing speed (keystrokes per minute)
\end{CodeBlock}
\end{tcolorbox}

\begin{tcolorbox}[breakable, title=Raw-Level Input Features, colback=white, colframe=black]
\begin{CodeBlock}
Given Data: In this context, the data was kept at the raw keystroke level. Unlike the word- and query-level settings, no aggregation or summary statistics are applied. Each keystroke event is preserved as an individual record and serialized into text, with the following five timing features recorded in milliseconds:

• Duration (raw)
• DD-time (raw)
• UD-time (raw)
• UU-time (raw)
• DU-time (raw)
\end{CodeBlock}
\end{tcolorbox}

\subsection{Few-shot examples}
We include few‑shot examples to provide context for expertise inference, enabling the model to ground its reasoning in realistic typing and query behaviors. Two configurations are considered: (i) \textit{within‑user} examples, which include queries and typing features from the same user, and (ii) \textit{cross‑user} examples, which are drawn from other users. The within‑user setting captures user‑specific typing dynamics and lexical preferences, while the cross‑user setting provides more diverse but potentially less aligned behavioral patterns. These examples help the model learn both individual behavioral cues and generalizable expertise‑related patterns. Here, each few‑shot example corresponds to the input representation shown in Appendix~\ref{appendix:feature_extraction} augmented with the associated expertise label, forming a set of $\{I, \hat{e}\}$ pairs. \cm{We sample these exemplars uniformly at random from eligible prior interactions using a fixed random seed of 0, excluding the current test query.}

\begin{tcolorbox}[breakable, title=Within-User Example Instructions , colback=white, colframe=black]
\begin{CodeBlock}
The following example records represent data from other queries and words entered by the same user in the past. Use them to understand the current user's query and typing patterns:
\end{CodeBlock}
\end{tcolorbox}

\begin{tcolorbox}[breakable, title=Cross-User Example Instructions , colback=white, colframe=black]
\begin{CodeBlock}
The following example records represent data from other queries and words entered by the other users in the past. 

Use them to understand the current user’s query and typing patterns:
\end{CodeBlock}
\end{tcolorbox}

\subsection{Semantic-only \& behavioral-only method prompt}
In addition to semantic‑behavioral representations, we also consider two ablation variants: (i) a \textit{semantic‑only} method, which uses only the raw query text as input without any keystroke features, and (ii) a \textit{behavioral‑only} method, which uses only the keystroke features without the semantic content of the query. In these settings, the input set was adjusted accordingly: the semantic‑only method received only the query text, while the behavioral‑only method received only the keystroke summary. Few‑shot examples were also constructed to match each input type, ensuring consistency between the example demonstrations and the target prompt.

\begin{tcolorbox}[breakable, title=Semantic-Only System Prompt, colback=white, colframe=black]
\begin{CodeBlock}
<core_identity>
You are an AI classifier that assigns a user's topic expertise level using ONLY the query text, with strict safeguards against overestimating expertise.
</core_identity>

<general_rules>
• NEVER use meta-phrases (e.g., "let me help you", "I can see that").
• ALWAYS be specific and grounded in observable evidence.
• Do NOT infer personal traits beyond what is required for classification.
• START IMMEDIATELY WITH THE ANSWER – ZERO INTRODUCTORY TEXT.
</general_rules>

<general_instructions>
Classify the user into exactly one category: Novice, Beginner, Intermediate, Proficient, Expert. Do not infer unobservable mental states.
</general_instructions>

<expertise_scale>
• Novice: A subject novice is a person who has little or no familiarity with the topic or domain of the conversation.
• Beginner: A subject beginner is someone who has little prior knowledge or experience in the topic or domain of the conversation.
• Intermediate: A subject intermediate is someone who has some basic knowledge or familiarity with the topic of the conversation, but not enough to be considered an expert or even proficient.
• Proficient: A subject proficient is someone who can apply relevant concepts and terminology from the conversation to different scenarios and problems.
• Expert: A subject expert is someone who has a deep and comprehensive understanding of the topic or field of the conversation and can use specialized terms and references to communicate their knowledge.
</expertise_scale>

<lemma technical_term_guidance>
Treat "technical term" as "domain-specific term with a specialized meaning in the given topic."
A query may contain technical terms even if the user is a novice (e.g., copied from class notes). Therefore, technical-term presence alone is weak evidence.
</lemma technical_term_guidance>

<classification_procedure>
Step 1) Compare the user's current query and the example query.
Step 2) Determine whether the metrics suggest novice, beginner, intermediate, proficient, or expert behavior patterns.
Step 3) Decide the most appropriate expertise level.
</classification_procedure>

<Examples>
{EXAMPLE_POOL_INSTRUCTION}
{EXAMPLES}
</Examples>

<output_format>
Output must be structured exactly in this format:

<classification>
  <level>One of: Novice, Beginner, Intermediate, Proficient, Expert</level>
  <reason>Brief explanation of the reasoning steps and observed indicators.</reason>
</classification>
</output_format>
\end{CodeBlock}
\end{tcolorbox}

\begin{tcolorbox}[breakable, title=Semantic-Only User Prompt, colback=white, colframe=black]
\begin{CodeBlock}
Given the following user input, determine the user’s expertise level.
Query: {USER_QUERY}
Expertise Level:
\end{CodeBlock}
\end{tcolorbox}

\begin{tcolorbox}[breakable, title=Behavioral-Only System Prompt, colback=white, colframe=black]
\begin{CodeBlock}
<core_identity>
You are an AI classifier that assigns a user's topic expertise level using BOTH the query text and keystroke dynamics, with strict safeguards against overestimating expertise.
</core_identity>

<general_rules>
• NEVER use meta-phrases (e.g., "let me help you", "I can see that").
• ALWAYS be specific and grounded in observable evidence.
• Do NOT infer personal traits beyond what is required for classification.
• START IMMEDIATELY WITH THE ANSWER – ZERO INTRODUCTORY TEXT.
</general_rules>

<general_instructions>
Classify the user into exactly one category: Novice, Beginner, Intermediate, Proficient, Expert. Do not infer unobservable mental states.
</general_instructions>

<expertise_scale>
• Novice: A subject novice is a person who has little or no familiarity with the topic or domain of the conversation.
• Beginner: A subject beginner is someone who has little prior knowledge or experience in the topic or domain of the conversation.
• Intermediate: A subject intermediate is someone who has some basic knowledge or familiarity with the topic of the conversation, but not enough to be considered an expert or even proficient.
• Proficient: A subject proficient is someone who can apply relevant concepts and terminology from the conversation to different scenarios and problems.
• Expert: A subject expert is someone who has a deep and comprehensive understanding of the topic or field of the conversation and can use specialized terms and references to communicate their knowledge.
</expertise_scale>

<keystroke_feature_definitions>
You are provided with a description of keystroke dynamics time-based features. Use this information to understand typing behavior and how different time intervals are defined.

Time-based features are defined for two consecutive keystroke events:
• Duration: The time that the user holds a key in the down position
• Down-down time (DD-time): The time between the press of a key and the press of the subsequent key
• Up-down time (UD-time): The time between the release of a key and the press of the subsequent key
• Up-up time (UU-time): The time between the release of a key and the release of the subsequent key
• Down-up time (DU-time): The time between the press of a key and the release of the subsequent key

Aggregate features:
• Typing speed (keystrokes per minute)
• Backspace count: The total number of Backspace key presses, representing correction activity.

{given_data}
</keystroke_feature_definitions>

<classification_procedure>
Step 1) Compare the user's current typing features to the example data to detect consistency or deviation.
Step 2) Evaluate typing speed and variability (standard deviation) as indicators of familiarity and fluency. And consider backspace frequency as an indicator of error correction and confidence.
Step 3) Determine whether the metrics suggest novice, beginner, intermediate, proficient, or expert behavior patterns.
</classification_procedure> 

<Examples>
{EXAMPLE_POOL_INSTRUCTION}
{EXAMPLES}
</Examples>

<output_format>
Output must be structured exactly in this format:

<classification>
  <level>One of: Novice, Beginner, Intermediate, Proficient, Expert</level>
  <reason>Brief explanation of the reasoning steps and observed indicators.</reason>
</classification>
</output_format>
\end{CodeBlock}
\end{tcolorbox}

\begin{tcolorbox}[breakable, title=Behavioral-Only User Prompt, colback=white, colframe=black]
\begin{CodeBlock}
Given the following user input, determine the user’s expertise level based on typing behavior.
Keystroke Summary: {BEHAVIORAL_FEATURE_SUMMARY}
Expertise Level:
\end{CodeBlock}
\end{tcolorbox}

\section{Expertise-conditioned response generation}
\label{appendix:expertise-conditioned-generation}
\subsection{Implementation details}
We used OpenAI’s \texttt{chatgpt-4o-latest} model~\cite{openai2024chatgpt4o} for all generation experiments, 
with \texttt{temperature}=1.0, \texttt{top\_p}=1.0, and a maximum output length of 2048 tokens to avoid truncation. This model was chosen because it closely reflects the behavior of widely available commercial chatbots, ensuring that our findings generalize to realistic user-LLM interaction settings.

\subsection{Expertise-conditioned response prompt}
To ensure unbiased ratings by preventing participants from recognizing whether the answer was personalized, an instruction explicitly stated that the response should never mention the user’s expertise.
\begin{tcolorbox}[breakable, title=System Prompt, colback=white, colframe=black]
\begin{CodeBlock}
Note: Never mention the [user's expertise] in the answer. Please write a response to the user's query according to [user's expertise] 
[user's expertise]: {USER_EXPERTISE}
[Novice] user has little to no prior knowledge of the topic and may ask vague or unfocused questions that sometimes stray from the main subject. Because they are encountering the material for the 
first time, they benefit most from simple, concrete explanations. Novices often struggle with misconceptions and confusion, requiring clear and intuitive feedback that emphasizes essential ideas while avoiding unnecessary complexity.
[Beginner] user has had some basic exposure to the topic and may recognize a few key terms, but lacks a deeper understanding of the concepts. They tend to seek definitions and foundational explanations, and they benefit from structured, step-by-step guidance. While they are beginning to engage with the subject, they still need help distinguishing between concepts and organizing information meaningfully.
[Intermediate] user is familiar with the major concepts and starts to ask more specific, application-oriented questions. They are developing analytical thinking skills and making initial attempts to relate new ideas to what they already know. While they understand commonly used terminology, they may still struggle with abstract or nuanced content and benefit from support that helps them connect and contextualize ideas.
[Proficient] user has a well-structured understanding of the domain and is capable of applying knowledge across different contexts. They ask integrative, evaluative, or comparative questions and use domain-specific language accurately. They value concise, conceptually rich explanations and are comfortable navigating complexity. At this level, clarity, precision, and coherence are particularly important.
[Expert] user possesses deep, specialized knowledge and engages with content at a high level of abstraction and theoretical depth. Their questions tend to be technically demanding, boundary-pushing, or aimed at synthesizing new insights. They are fluent in the language of the domain and expect dense, 
information-rich responses that prioritize conceptual rigor, accuracy, and originality.
\end{CodeBlock}
\end{tcolorbox}

\begin{tcolorbox}[breakable, title=User Prompt, colback=white, colframe=black]
\begin{CodeBlock}
{USER_QUERY}
\end{CodeBlock}
\end{tcolorbox}

\section{Feature Extraction}
\label{appendix:feature_extraction}
To support expertise inference from keystroke dynamics, we construct explicit semantic-behavioral input feature representations that are directly fed into our expertise inference prompt as user prompt (Appendix~\ref{appendix:expertise_inference}). These inputs integrate (i) the raw query text, capturing semantic indicators of the user’s domain knowledge, and (ii) numerical feature vectors derived from keystroke dynamics, capturing behavioral cues of typing fluency and confidence. Below, we provide concrete examples for each feature format to illustrate the exact values passed to the inference model.


\subsection{Word-level input examples}
Word-level features capture fine-grained typing behavior by segmenting the query into individual words and computing a feature vector for each. For every word segment, we compute five timing-based metrics---key hold duration, Down-Down (DD), Up-Down (UD), Up-Up (UU), and Down-Up (DU) intervals---and aggregate each by its mean and standard deviation. We additionally record two global control signals: typing speed (keystrokes per minute) and the number of backspace keystrokes used during that word. These features are paired with the corresponding word tokens to align behavioral patterns with semantic context.

\begin{tcolorbox}[breakable, colback=white, colframe=black, title=Word-level Input Feature Example]
\begin{CodeBlock}
Query: Explain the symmetry elements of molecules. 
Keystroke Summary:
[Word] <dummy> 
[Typing Features] 
• Duration: mean=65.00 / std=57.66 
• DD-time: mean=127.67 / std=110.75 
• UD-time: mean=62.67 / std=57.50 
• UU-time: mean=117.00 / std=64.28 
• DU-time: mean=182.00 / std=95.54 
• Total number of Backspace keys: 3 
• Typing speed (keystrokes per minute): 972.97 

[Word] Explain 
[Typing Features] 
• Duration: mean=119.07 / std=125.65 
• DD-time: mean=4109.47 / std=21507.16 
• UD-time: mean=3990.40 / std=21506.09 
• UU-time: mean=4246.33 / std=21489.42 
• DU-time: mean=4365.40 / std=21494.63 
• Total number of Backspace keys: 7 
• Typing speed (keystrokes per minute): 168.15 

[Word] the 
[Typing Features] 
• Duration: mean=90.67 / std=28.26 
• DD-time: mean=131.83 / std=49.22 
• UD-time: mean=41.17 / std=61.52 
• UU-time: mean=124.17 / std=72.35 
• DU-time: mean=214.83 / std=59.62 
• Total number of Backspace keys: 0 
• Typing speed (keystrokes per minute): 516.50 

...

[Word] molecules. 
[Typing Features] 
• Duration: mean=101.92 / std=27.61 
• DD-time: mean=91.69 / std=35.30 
• UD-time: mean=-10.23 / std=56.46 
• UU-time: mean=97.23 / std=65.89 
• DU-time: mean=199.15 / std=49.47 
• Total number of Backspace keys: 0 
• Typing speed (keystrokes per minute): 687.22
\end{CodeBlock}
\end{tcolorbox}

\subsection{Query-level input examples}
Query-level features summarize the entire query as a single behavioral profile. Instead of computing features for each word separately, all timing intervals, typing speed, and backspace counts are aggregated across the full query. 

\begin{tcolorbox}[breakable, colback=white, colframe=black, title=Query-level Input Feature Example]
\begin{CodeBlock}
Query: Explain the symmetry elements of molecules. 

Keystroke Summary:
Duration (mean/std): 109.42 / 88.84 
DD-time (mean/std): 1994.44 / 14728.94 
UD-time (mean/std): 1885.02 / 14727.02 
UU-time (mean/std): 2057.08 / 14726.80 
DU-time (mean/std): 2166.50 / 14731.55 
Typing speed (kpm): 346.32 
Backspace count: 10    
\end{CodeBlock}
\end{tcolorbox}

\subsection{Raw-level input examples}
Raw-level features retain keystroke dynamics for each key event rather than aggregating across the entire query. For every key pressed, we record five timing metrics in milliseconds: key hold duration (\texttt{Duration\_key1}), time between consecutive key presses (\texttt{DD-time}), time between a key release and the next key press (\texttt{UD-time}), time between consecutive key releases (\texttt{UU-time}), and time between the press of one key and the release of the next (\texttt{DU-time}). 

\begin{tcolorbox}[breakable, colback=white, colframe=black, title=Raw-level Input Feature Example]
\begin{CodeBlock}
Query: Explain the symmetry elements of molecules.

Raw Keystroke Features (excerpt):
Key: E, Duration_key1=120, DD_time=95, UD_time=102, UU_time=110, DU_time=108 
Key: x, Duration_key1=98,  DD_time=87, UD_time=93,  UU_time=101, DU_time=97  
Key: p, Duration_key1=105, DD_time=92, UD_time=97,  UU_time=104, DU_time=99  
Key: l, Duration_key1=130, DD_time=100, UD_time=106, UU_time=113, DU_time=109 
Key: a, Duration_key1=90,  DD_time=85, UD_time=88,  UU_time=96,  DU_time=91  
... (remaining keys omitted for brevity)
\end{CodeBlock}
\end{tcolorbox}

\section{User Study: Task and Interface Information}
\label{appendix:userstudy}
Before starting the study, the protocol was approved by the local Institutional Review Board (IRB). All participants provided informed consent, approved by the local Institutional Review Board (IRB), which permitted the collection and analysis of their interaction data. All users were anonymized, and all data were stored and used strictly for research purposes. All participants were compensated 22 USD, and to enhance engagement, the top five participants, based on question quality and completion, received an additional 15 USD reward. 

\subsection{User study interface}
\label{appendix:userstudy-interface}
\begin{figure*}[t]
    \centering
    \includegraphics[width=\linewidth]{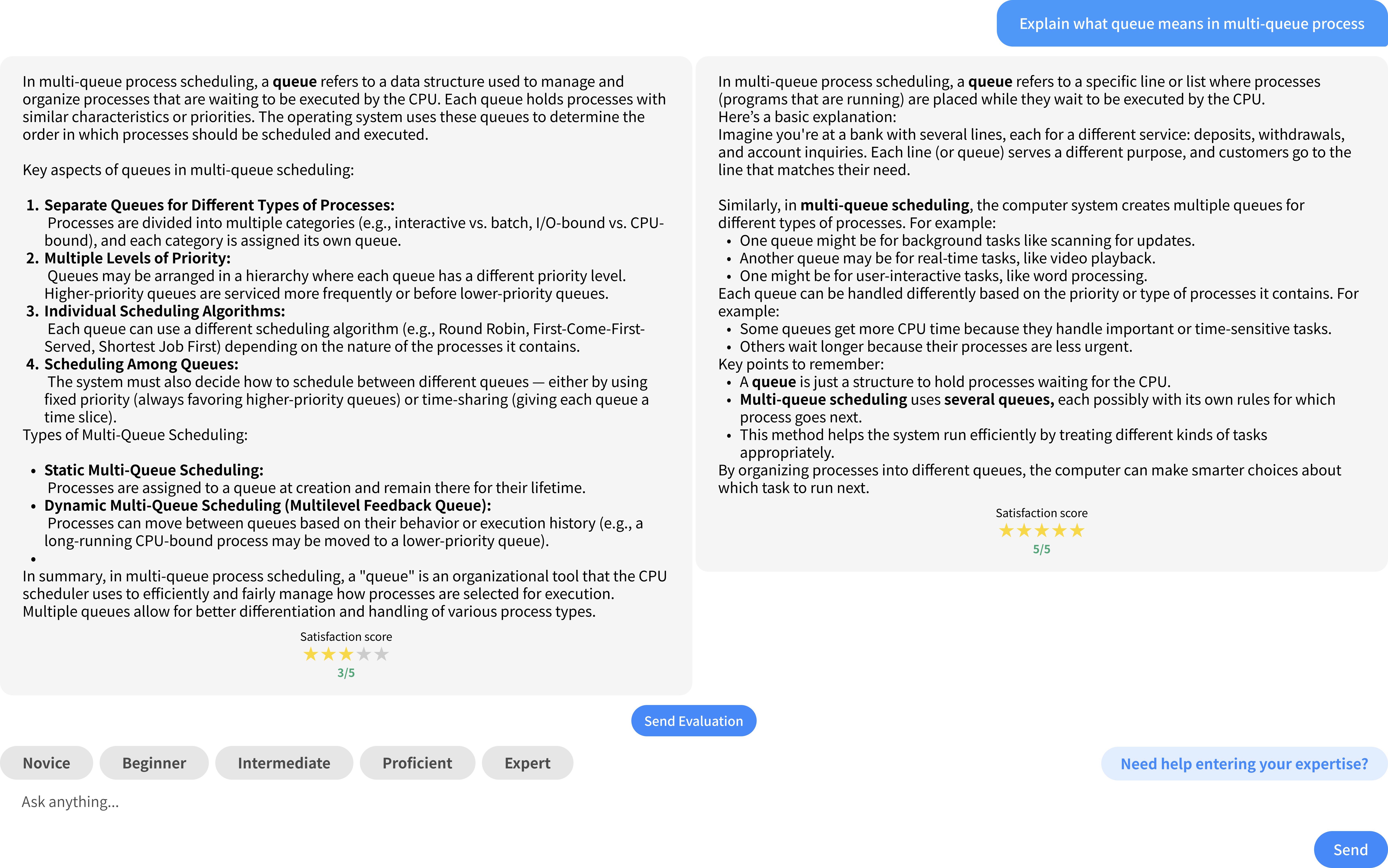}
    \caption{Custom web-based user study interface for capturing detailed user--LLM interaction data. Participants entered queries, self-reported their expertise level using a five-point scale (Novice$\sim$Expert), and rated response satisfaction on a five-point Likert scale after each response.}
    \label{fig:interface}
\end{figure*}

We implemented a custom web-based interface to capture detailed user-LLM interaction data while maintaining experimental control and minimizing distractions. Each session, which lasted approximately 2 to 2.5 hours in total, was conducted on a lab-provided computer under the supervision of a study operator. \cm{For each query, participants were presented with two responses and asked to comparatively evaluate them. To reduce presentation-order bias, the responses were displayed in randomized left–right order. Participants were blind to the condition and were not informed which response came from the baseline system or the expertise-conditioned system. To further ensure a fair comparison, all responses were generated using the same decoding settings across conditions, thereby controlling for verbosity and other generation-related differences. }The interface and study flow were structured as follows:

\begin{itemize}
    \item \textbf{Task Execution.} Each participant performed three distinct information-seeking tasks, each lasting 25 minutes, using the chatbot interface. Participants were given detailed written instructions for each task on printed handouts. The chatbot’s conversational state was reset between tasks to avoid carryover effects. In the experiment, participants completed three domain-specific tasks, with the task order balanced using a partial counterbalancing method.
    \item \textbf{Query Submission and Feedback.} Before submitting each query, participants rated their familiarity with the topic using a five-level scale (Novice-Expert). A tooltip-based help system clarified the meaning of each level. After receiving the LLM’s response, participants rated their satisfaction on a mandatory five-point Likert scale (1: very unsatisfied, 5: very satisfied) using a star-rating widget.
    \item \textbf{Interaction Logging.} The interface automatically logged all user interactions, including timestamps, queries, self-reported expertise, model responses, and satisfaction ratings for subsequent analysis.
    \item \textbf{Experimental Controls and Compliance.} To ensure engagement and control for external aids, copy-paste functionality was disabled, and participants were required to submit a minimum of ten queries per task. The chatbot was stateless across tasks and did not retain prior interactions, which was explicitly communicated to participants to discourage meta-queries (e.g., “continue from earlier”). 
\end{itemize}

After completing the three tasks, participants filled out a brief interaction survey (approximately two minutes) regarding perceived difficulty and usability, followed by a nine-item multiple-choice quiz assessing knowledge acquisition based on the investigated topics.

\subsection{User study tasks}
\label{appendix:userstudy-tasks}

Participants completed three distinct information-seeking tasks, each designed to reflect a different academic domain and to evaluate the LLM’s effectiveness in supporting domain-specific knowledge acquisition. Each task lasted 25 minutes and was accompanied by detailed written instructions displayed within the study interface. The tasks were as follows:

\begin{itemize}
    \item \textbf{Chemistry} Participants investigated the principles of molecular symmetry and the classification of molecules into point groups. Subtasks included: (i) identifying symmetry elements such as rotation axes, centers of inversion, and mirror planes; (ii) explaining the criteria for point group classification; and (iii) discussing correlations between point groups and spectroscopic properties, including infrared (IR) and Raman activity.
    \item \textbf{Computer Science} 
Participants explored two core process scheduling algorithms used in operating systems: priority scheduling and multilevel queue scheduling. Subtasks included: 
(i) analyzing the design and structure of priority scheduling; 
(ii) investigating the implementation of multilevel queue scheduling; and 
(iii) examining supporting mechanisms such as data structure design, time allocation methods, and priority handling logic.
    \item \textbf{Business} Participants studied key accounting issues encountered during the preparation of consolidated financial statements. Focus areas included: (i) elimination of investments in subsidiaries against equity; (ii) recognition and treatment of goodwill or bargain purchase gains; and (iii) accounting for non-controlling interests.
\end{itemize}

\subsection{Participant demographics}
\label{appendix:participants}

Table~\ref{tab:participant-demographics} summarizes the demographic and academic background information of the study participants. The collected data include age (in years), gender, nationality, major or primary academic field, and current academic year.

\begin{table}[t]
\color{revcolor}
\captionsetup{labelfont={color=revcolor}, textfont={color=revcolor}}
\centering
\caption{Participant demographics and academic background.}
\label{tab:participant-demographics}
\resizebox{\linewidth}{!}{%
\begin{tabular}{rlll}
\toprule
\textbf{Age} & \textbf{Gender} & \textbf{Major (Double or Minor)} & \textbf{Academic Year} \\
\midrule
24 & Male   & Artificial Intelligence                   & Graduate (Master’s) \\

20 & Female & Business Administration                   & 3rd year \\
21 & Female & Business Administration                   & 3rd year \\
22 & Male   & Business Administration                   & 3rd year \\
26 & Male   & Business Administration                   & 4th year \\

22 & Female & Chemical Engineering                      & Graduate (Master’s) \\
24 & Male   & Chemical Engineering                      & 4th year \\
24 & Female & Chemical and Biological Engineering       & 4th year \\

20 & Female & Computer Science                          & 3rd year \\
21 & Female & Computer Science                          & 3rd year \\
21 & Male   & Computer Science                          & 3rd year \\
23 & Male   & Computer Science                          & 3rd year \\
23 & Male   & Computer Science                          & 3rd year \\
24 & Male   & Computer Science                          & 3rd year \\
22 & Male   & Computer Science                          & 4th year \\
23 & Female & Computer Science                          & 4th year \\
24 & Male   & Computer Science                          & 4th year \\
23 & Female & Computer Science                          & Graduate (Master’s) \\
24 & Male   & Computer Science                          & Graduate (Master’s) \\
25 & Female & Computer Science                          & Graduate (Master’s) \\

21 & Female & Energy Chemical Engineering               & 3rd year \\
23 & Male   & Energy Chemical Engineering               & 3rd year \\
23 & Male   & Energy Chemical Engineering               & 3rd year \\
23 & Male   & Energy Chemical Engineering               & 3rd year \\
22 & Female & Energy Chemical Engineering               & 4th year \\
23 & Female & Energy Chemical Engineering               & 4th year \\
24 & Male   & Energy Chemical Engineering               & 4th year \\
25 & Female & Energy Chemical Engineering               & Graduate (Master’s) \\
24 & Female & Energy Chemical Engineering               & Graduate (Master’s) \\

22 & Female & Industrial Engineering (Management Sci.)  & 4th year \\
22 & Female & Industrial Engineering (Management Sci.)  & 4th year \\
24 & Male   & Industrial Engineering (Management Sci.)  & 4th year \\

21 & Female & Management Science                        & 3rd year \\
23 & Male   & Management Science                        & 3rd year \\
27 & Male   & Management Engineering                    & Graduate (Master’s) \\

22 & Male   & Economics                                 & 3rd year \\
25 & Female & Economics                                 & 4th year \\

24 & Female & Chemistry                                 & 4th year \\
24 & Female & Materials Science and Engineering         & Graduate (Master’s) \\
26 & Male   & Industrial Design (Minor: Management)     & 4th year \\
\bottomrule
\end{tabular}
}
\end{table}

\subsection{Collected query expertise results}

\begin{table}[t]
\color{revcolor}
\captionsetup{labelfont={color=revcolor}, textfont={color=revcolor}}
\centering
\caption{Distribution of collected queries across self‑reported expertise levels for each domain, illustrating the coverage of expertise levels in the dataset.}
\label{tab:Nqeury-by-expertise}
\resizebox{\linewidth}{!}{%
\begin{tabular}{lccc}
\toprule
\textbf{Expertise Level} & \textbf{Chemistry} & \textbf{Computer Science} & \textbf{Business} \\
\midrule
Novice            & 206 & 188 & 257 \\
Beginner          & 86 & 72 & 87 \\
Intermediate      & 95 & 69 & 45 \\
Proficient        & 32 & 58 & 30 \\
Expert            & 4 & 31 & 10 \\
\bottomrule
\end{tabular}}
\end{table}

For each collected query, we recorded the query text, the user’s self‑reported expertise level, two LLM responses (baseline and expertise‑conditioned), and satisfaction ratings, along with raw keystroke logs capturing key press and release events with timestamps. In total, the dataset comprised 1270 queries spanning Chemistry, Computer Science, and Business, covering a range of expertise levels (651 Novice, 245 Beginner, 209 Intermediate, 120 Proficient, and 45 Expert queries). Table~\ref{tab:Nqeury-by-expertise} shows how queries were distributed across expertise levels within each domain, indicating that most queries came from novice users, with decreasing numbers at higher expertise levels and some variation between domains.


\subsection{Survey results}
\label{appendix:survey-results}
\subsubsection{Quiz performance by self-assessed expertise}


To analyze participants' objective knowledge across domain-specific tasks, we examined quiz scores (out of 9) stratified by self-assessed expertise level for each task. Participants provided their overall expertise per task on a scale: Novice, Beginner, Intermediate, Proficient, and Expert. Given the relatively few Expert self-assessments, we combined Proficient and Expert into a higher proficiency group for analysis.

Overall, the results show a strong positive association between self-assessed domain expertise and objective knowledge, as measured by quiz scores. Across all three tasks, mean quiz performance increased monotonically with expertise level. For example, participants identifying as Proficient or Expert consistently scored above 7.0, whereas Novices averaged between 5.1 and 5.8 points. Notably, the magnitude of expertise-related differences varied by domain. The largest score gap was observed in Chemistry (Task 1), where Proficient/Expert participants averaged 7.3 compared to 5.1 among Novices (a 2.2-point difference). Computer Science (Task 2) exhibited the highest overall scores, with even Novices achieving a mean of 5.7, while Business (Task 3) showed intermediate patterns. Although the sample sizes for higher proficiency levels were small, the trend remains consistent across tasks. These results indicate that self-perceived expertise reliably predicts task-specific knowledge, while also revealing substantial domain-specific variability in baseline knowledge levels among participants.

\begin{table*}[t]
\color{revcolor}
\captionsetup{labelfont={color=revcolor}, textfont={color=revcolor}}
\centering
\footnotesize  
\caption{Mapping between full survey questions, their scale type, and the abbreviated labels used in Table~\ref{tab:subjective-ratings-expertise-updated}.}
\label{tab:appendix-question-mapping}
\begin{tabular}{p{7cm}p{3cm}p{5cm}}
\toprule
\textbf{Survey Question} & \textbf{Scale (Range)} & \textbf{Label in Results Table} \\
\midrule
My overall expertise for this task was sufficient. & Likert (1--7) & Task appropriateness \\
The difficulty of this task was appropriate relative to my ability. & Likert (1--7) & Task challenge \\
This task was an appropriate challenge for me. & Likert (1--7) & Task challenge \\
I felt immersed while performing this task. & Likert (1--7) & Immersion \\
The chatbot provided effective assistance in completing this task. & Likert (1--7) & LLM Helpfulness \\
The information provided by the chatbot was appropriate for my level of expertise. & Likert (1--7) & LLM Relevance \\
How mentally demanding was the task? & NASA-TLX (0--20) & Mental demand \\
How physically demanding was the task? & NASA-TLX (0--20) & Physical demand \\
How hurried or rushed was the pace of the task? & NASA-TLX (0--20) & Pace/hurriedness \\
How successful were you in accomplishing what you were asked to do? & NASA-TLX (0--20) & Task success \\
How hard did you have to work to achieve your level of performance? & NASA-TLX (0--20) & Effort needed \\
How insecure, discouraged, irritated, stressed, or annoyed did you feel during the task? & NASA-TLX (0--20) & Negative affect \\
\bottomrule
\end{tabular}
\end{table*}

\begin{table*}[t]
\color{revcolor}
\captionsetup{labelfont={color=revcolor}, textfont={color=revcolor}}
\centering
\caption{Summary of Subjective Ratings by Self-Assessed Expertise}
\label{tab:subjective-ratings-expertise-updated}
\resizebox{0.7\linewidth}{!}{%
\begin{tabular}{lrrrr}
\toprule
\textbf{Measure} & \textbf{Novice} & \textbf{Beginner} & \textbf{Intermediate} & \textbf{Proficient/Expert} \\
\midrule
Task appropriateness (1--7)   & 3.5 & 5.0 & 5.4 & 6.4 \\
Task challenge (1--7)         & 5.0 & 5.2 & 4.8 & 4.2 \\
Immersion (1--7)              & 4.9 & 5.7 & 5.3 & 5.8 \\
LLM Helpfulness (1--7)        & 5.8 & 5.8 & 5.9 & 6.4 \\
LLM Relevance (1--7)          & 4.2 & 5.8 & 5.1 & 3.2 \\
Mental demand (0--20)         & 11.6 & 7.8 & 6.9 & 6.2 \\
Physical demand (0--20)       & 4.2 & 3.0 & 3.4 & 3.2 \\
Pace/hurriedness (0--20)      & 6.9 & 5.9 & 6.5 & 5.9 \\
Task success (0--20)          & 9.2 & 7.8 & 7.8 & 9.5 \\
Effort needed (0--20)         & 12.2 & 12.2 & 11.1 & 13.8 \\
Negative affect (0--20)       & 7.8 & 4.7 & 4.8 & 3.1 \\
\bottomrule
\end{tabular}}
\end{table*}

\subsubsection{Task experience and cognitive load by self-assessed expertise}
\label{appendix:survey-results-cognitive-load}

Participants rated multiple aspects of their task experience using a combination of standardized and custom measures. Table~\ref{tab:appendix-question-mapping} details the full survey questions, their scale types, and the abbreviated labels used in reporting, while Table~\ref{tab:subjective-ratings-expertise-updated} summarizes mean ratings for each measure stratified by self-assessed expertise group. The measures included perceptions of task appropriateness, difficulty, and immersion, as well as the perceived helpfulness and relevance of LLM assistance, all rated on 7-point Likert scales. Additionally, six NASA Task Load Index (NASA-TLX) subscales were administered, capturing cognitive and emotional workload dimensions (e.g., mental demand, effort, and negative affect) on 0–20 scales.

Consistent with expectations, self-assessed expertise was systematically associated with participants’ subjective task experiences (Table~\ref{tab:subjective-ratings-expertise-updated}). Participants with higher expertise generally perceived the tasks as better matched to their abilities and reported lower levels of cognitive workload. In particular, Novice participants reported higher mental demand than Proficient/Expert participants (11.6 vs.\ 6.2). Negative affect (e.g., stress and frustration) likewise decreased with increasing expertise (7.8 vs.\ 3.1), indicating a more positive emotional experience among more experienced users.
Perceived task success showed a modest increase with expertise (9.2 for Novices vs.\ 9.5 for Proficient/Expert participants). In contrast, effort ratings were comparable across expertise levels, with Proficient/Expert participants reporting slightly higher effort (13.8) than Novices (12.2), despite lower reported mental demand.
Ratings of LLM helpfulness and relevance showed relatively small differences across expertise levels(5.8–6.4), suggesting broadly consistent perceptions of system support. Overall, these results highlight systematic expertise-related differences in cognitive workload and affect, illustrating how participants’ perceived proficiency shaped their subjective task experience.
\cm{
\subsection{Follow-up Evaluation}
\label{sec:followup}
For the follow-up study, we recontacted participants from the original study who had agreed to be contacted again for future research. Among those we were able to reach, 16 returning participants completed the follow-up evaluation and were included in the analysis. Participants received an additional compensation of 15 USD for completing the study.

In this follow-up evaluation, participants rated responses to queries they had originally written in the main study. For each query, they were shown responses from three conditions: (1) a baseline response without expertise conditioning, (2) a response conditioned on self-reported expertise, and (3) a response conditioned on ~\system{}-predicted expertise. The prompts used for expertise-conditioned generation were identical to those described in Appendix ~\ref{appendix:expertise-conditioned-generation}, with the predicted-expertise condition using the expertise level inferred by ~\system{} instead of self-reported expertise. The order of the conditions was randomized, and the identities of the conditions were blinded to participants. Participants rated each response on a 5-point satisfaction scale. This design allowed us to assess whether the benefits of expertise-conditioned generation persisted when the expertise signal was inferred by the model rather than explicitly provided by the user.}

\begin{figure}[h]
    \centering
    \includegraphics[width=\linewidth]{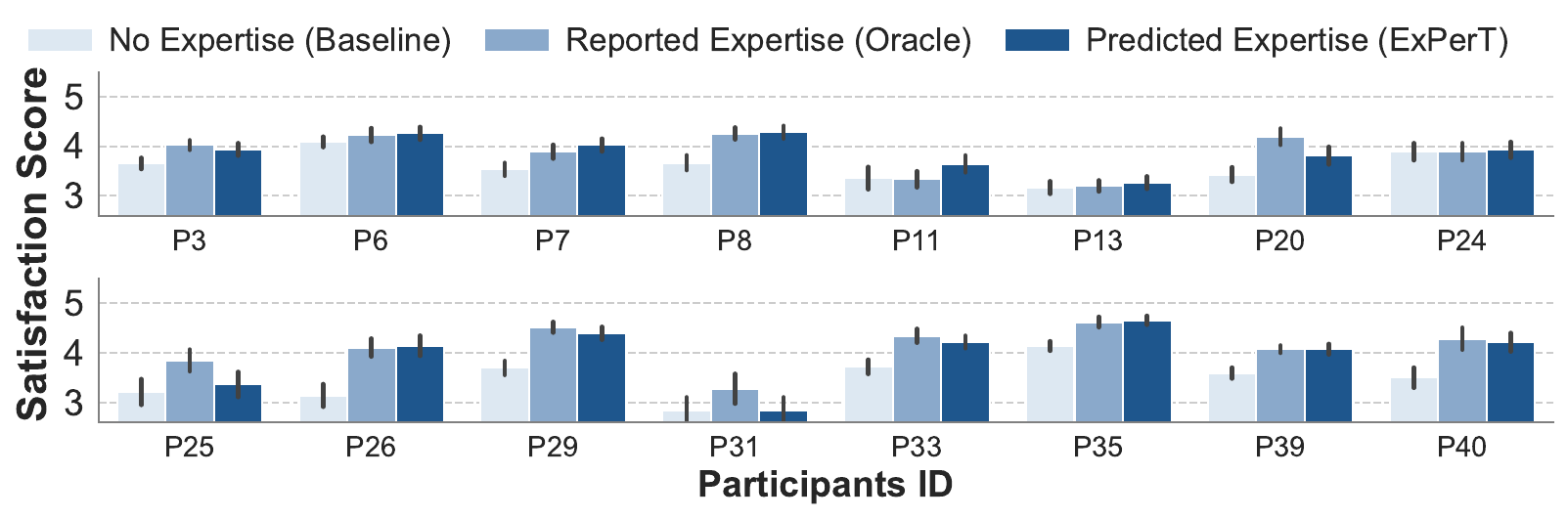}
    \caption{User satisfaction scores for each participant across the without expertise, self-reported expertise, and predicted expertise conditions.}
    \label{fig:follow-up-study}
\end{figure}

As shown in Figure ~\ref{fig:follow-up-study}, most participants preferred responses that incorporated expertise information over those that did not. At the individual level, both the reported-expertise and predicted-expertise conditions generally received higher satisfaction scores than the no-expertise condition, indicating that expertise-aware responses were more positively evaluated overall. While some participant-level variation remained, the advantage of the expertise-based conditions was consistently observed across a broad set of users rather than concentrated in only a few cases. Furthermore, the reported-expertise and predicted-expertise conditions were competitive, with many participants assigning similarly high ratings to both. This pattern suggests that predicted expertise can approximate the benefit of reported expertise in supporting user satisfaction. Taken together, these findings indicate that incorporating expertise information is beneficial at the participant level, and that inferred expertise may provide a practical alternative when self-reported expertise is unavailable.

\section{Baselines for expertise inference}
\label{appendix:baselines}
To benchmark our proposed approach, we implemented \textbf{random guess baseline} and two baseline expertise classification systems following prior work on persona-aware LLM prompting \cite{hu-collier-2024-quantifying,palta2025speakingrightlanguageimpact}: a \textbf{persona-based baseline} and a \textbf{session-based baseline}. We additionally report a \textbf{conventional supervised classifier baseline} for completeness, although it is intended primarily as a supplementary reference point.

\subsection{Baseline Details}
\paragraph{Random guess baseline}
For reference, we compute the expected performance of a random predictor that samples labels uniformly from the five-point ordinal expertise scale $\mathcal{Y} = \{1, 2, 3, 4, 5\}$. Assuming that the true label $Y$ is fixed and the prediction $\hat{Y}$ is drawn uniformly at random from $\mathcal{Y}$, the expected mean squared error (MSE) and mean absolute error (MAE) are given by
\[
\mathbb{E}\!\left[(Y - \hat{Y})^2\right]
= \frac{1}{5} \sum_{y=1}^{5}\frac{1}{5}\sum_{\hat{y}=1}^{5} (y-\hat{y})^2
= 4.0,
\]
\[
\mathbb{E}\!\left[|Y - \hat{Y}|\right]
= \frac{1}{5} \sum_{y=1}^{5}\frac{1}{5}\sum_{\hat{y}=1}^{5} |y-\hat{y}|
= 1.6.
\]
These values represent a random guess baseline, which any model leveraging semantic or behavioral information should outperform.

\paragraph{Persona-based baseline}
We implement a persona-aware classifier that predicts user expertise using both query text and structured persona metadata 
(age, gender, nationality, native and foreign languages, academic major, study year, and self-reported LLM usage). 
For each query, the persona attributes are presented in the system prompt and the query text in the user prompt 
(see Box~\ref{box:persona-prompt}). The model outputs one of five ordered labels (Novice, Beginner, Intermediate, Proficient, Expert) along with a concise justification.

\begin{tcolorbox}[breakable, title=Persona-Based System Prompt, colback=white, colframe=black, width=\linewidth]
\label{box:persona-prompt}
\begin{CodeBlock}
<core_identity>
You are an AI classifier that determines a user's expertise level.
</core_identity>

<persona_context>
Persona:
- Age: {age}
- Gender: {gender}
- Nationality: {nationality}
- Native language: {native_language}
- First foreign language: {first_foreign_language}
- Other languages: {other_languages}
- Major: {major}
- Year: {year}
- LLM usage frequency: {llm_usage}
</persona_context>

<general_rules>
• NEVER use meta-phrases (e.g., "let me help you", "I can see that").
• ALWAYS be specific and grounded in observable evidence.
• Do NOT infer personal traits beyond what is required for classification.
• START IMMEDIATELY WITH THE ANSWER – ZERO INTRODUCTORY TEXT.
<general_rules>

<general_instructions>
Classify the user into exactly one category: Novice, Beginner, Intermediate, Proficient, Expert. Do not infer unobservable mental states.
</general_instructions>

<expertise_scale>
• Novice: A subject novice is a person who has little or no familiarity with the topic or domain of the conversation.
• Beginner: A subject beginner is someone who has little prior knowledge or experience in the topic or domain of the conversation.
• Intermediate: A subject intermediate is someone who has some basic knowledge or familiarity with the topic of the conversation, but not enough to be considered an expert or even proficient.
• Proficient: A subject proficient is someone who can apply relevant concepts and terminology from the conversation to different scenarios and problems.
• Expert: A subject expert is someone who has a deep and comprehensive understanding of the topic or field of the conversation and can use specialized terms and references to communicate their knowledge.
</expertise_scale>

<classification_procedure>
Before outputting the final classification, follow this reasoning process internally:
Step 1) Consider the full persona of the user (age, gender, nationality, native and foreign languages, major, academic year, and frequency of LLM usage).
Step 2) Evaluate how this persona might correlate with their knowledge depth and experience relevant to the query topic.
Step 3) Compare the style, complexity, and focus of the query against typical expertise-level patterns.
Step 4) Combine the persona context and query characteristics to infer the most likely expertise level.
Step 5) Decide the most appropriate expertise level.
</classification_procedure>

<output_format>
Output must be structured exactly in this format:

<classification>
  <level>One of: Novice, Beginner, Intermediate, Proficient, Expert</level>
  <reason>Brief explanation of the reasoning steps and observed indicators.</reason>
</classification>
</output_format>
\end{CodeBlock}
\end{tcolorbox}

\begin{tcolorbox}[breakable, title=Persona-Based User Prompt, colback=white, colframe=black, width=\linewidth]
\begin{CodeBlock}
Given the following user input and persona information, determine the user’s expertise level.
Query: {USER_QUERY}
Expertise Level:
\end{CodeBlock}
\end{tcolorbox}

All experiments used OpenAI’s \texttt{gpt-4.1} with \texttt{temperature}=0.0, \texttt{top\_p}=1.0, and a maximum output length of 512 tokens to ensure deterministic outputs without truncation. 

\paragraph{Session-based baseline}
We implement a session-level classifier that predicts user expertise using only the linguistic content of all queries entered by a participant for a given task. Each session (grouped by \texttt{user\_id} and \texttt{task}) is concatenated into a single text block and classified using GPT-4.1 with \texttt{temperature}=0.0, \texttt{top\_p}=1.0, and a 512-token output cap to ensure deterministic responses. We adopt the session-level prompt template from Palta et al.~\cite{palta2025speakingrightlanguageimpact} without modification, where the model infers an overall expertise label (Novice-Expert) based solely on aggregated query text.

Predictions are evaluated against ground-truth expertise labels obtained by aggregating participants’ self-assessed query-level ratings within each session.

\paragraph{IDL baseline}
We implement an IDL-based baseline that adapts the core principles of In-Dialogue Learning (IDL)~\cite{cheng2024dialogues} to expertise inference, where stable user characteristics are induced from accumulated dialogue history without relying on predefined profiles. The baseline uses \texttt{LLaMA-2-7B-Chat} as the backbone model and operates on query histories spanning multiple turns and multiple dialogue sessions for each user-task pair.
For each user, dialogue sessions are embedded using a sentence-level encoder and clustered via $k$-means to identify persona-relevant session groups. Within each cluster, reference sessions are dynamically reordered by minimizing inter-session conversational edit distance, producing a coherent dialogue sequence. The model is then trained using a Mutual Supervised Learning (MSL) objective following the IDL framework, with LoRA-based fine-tuning, and subsequently aligned via DPOC-based preference optimization to mitigate persona inconsistency and hallucination. At inference time, the LoRA- and DPOC-aligned model is provided with the concatenated dialogue history and prompted to infer the user’s overall expertise level on a five-point ordinal scale (Novice-Expert).

\paragraph{AI Persona baseline}
We implement an AI Persona baseline following the life-long semantic personalization framework proposed in AI Persona~\cite{wang2024ai}. This approach models user characteristics as an explicit persona state that is incrementally updated from accumulated interaction history, under the assumption that stable user attributes can be inferred through repeated dialogue with a language model.
Concretely, for each user, we maintain a structured persona representation initialized from demographic and background information. Persona updates are performed by aggregating interaction histories over multiple sessions and prompting GPT-4.1 to revise the persona fields accordingly. Following the original AI Persona protocol, persona updates are triggered every $k=3$ sessions, where all query---response pairs within the update window are concatenated and provided to the model for persona revision.
For expertise prediction, the current persona state is combined with the user’s query and provided to GPT-4.1, which is prompted to classify the user’s expertise level on a five-point ordinal scale (Novice to Expert). All persona updates and expertise predictions are performed using GPT-4.1 with temperature set to 0.0, top\_p set to 1.0, and a maximum output length of 512 tokens to ensure deterministic behavior. This baseline relies exclusively on semantic information encoded in dialogue content and persona representations, and does not incorporate query-level behavioral signals or fine-grained temporal features.

\cm{\subsection{Conventional Supervised Classifier Baselines}
To further examine whether ~\system{}'s gains stem from its in-context semantic-behavioral inference rather than from multimodal input availability alone, we additionally implemented three conventional supervised classifier baselines using the same query text and keystroke-derived inputs: (i) BiLSTM(text) + FFN(keystroke\_flat) ~\cite{zadeh-etal-2017-tensor, liu-etal-2018-efficient-low}, (ii) Word2Vec(text) + FFN(keystroke\_flat) ~\cite{tong-etal-2017-combating}, and (iii) BiLSTM(text) + BiLSTM(keystroke\_seq) ~\cite{zadeh2018memory} with attention pooling and late fusion (concatenation followed by an MLP). These baselines are included as additional supervised reference points.

\paragraph{Inputs and architectures.}
For all baselines, the text modality is constructed from the query text, and the behavioral modality is constructed from the same keystroke-derived signals used in our main task. The BiLSTM(text) + FFN(keystroke\_flat) baseline encodes the query text with a BiLSTM and flattened keystroke features with a feed-forward network, followed by fusion for prediction. The Word2Vec(text) + FFN(keystroke\_flat) baseline represents the query text using averaged Word2Vec embeddings and combines it with an FFN encoding of flattened keystroke features. The BiLSTM(text) + BiLSTM(keystroke\_seq) baseline encodes the query text and keystroke sequence with separate BiLSTMs, applies attention pooling, and performs late fusion via concatenation followed by an MLP. In all cases, the final prediction target is our 5-level expertise label.

\paragraph{Evaluation setting.}
We evaluate these supervised baselines using the \textit{individual} split, where each user's samples are divided into train/validation/test sets with a 70/15/15 ratio. Since this setting is the closest supervised counterpart to our within-user evaluation, we compare these results against ~\system{} under the within-user 10-shot setting.

\paragraph{Results.}
Table~\ref{tab:additional_baselines} reports the results in terms of mean absolute error (MAE) and mean squared error (MSE), where lower values indicate better performance. Among the supervised baselines, BiLSTM + FFN achieves the best MAE, while Word2Vec + FFN achieves the best MSE. However, all three supervised baselines remain substantially behind ~\system{}, indicating that ~\system{}'s gains are not explained solely by access to both semantic and behavioral inputs, but are associated with its in-context LLM-based inference mechanism.

\begin{table}[t]
\centering
\small
\begin{tabular}{lcc}
\toprule
\textbf{Model} & \textbf{MAE} $\downarrow$ & \textbf{MSE} $\downarrow$ \\
\midrule
BiLSTM + FFN & 0.865 & 2.032 \\
Word2Vec + FFN & 0.902 & 1.894 \\
BiLSTM + BiLSTM & 1.026 & 2.216 \\
~\system{} & \textbf{0.398} & \textbf{0.698} \\
\bottomrule
\end{tabular}
\caption{Comparison between conventional supervised classifier baselines and ~\system{}. Lower MAE and MSE indicate better performance.}
\label{tab:additional_baselines}
\end{table}

\section{Per-class metrics}
We additionally report per-class F1 scores with 95\% bootstrap confidence intervals (1,000 resamples with replacement) in Table~\ref{tab:per-class-f1}. The overall macro-F1 is 0.6102 with 95\% CI [0.5682, 0.6479].

Performance is highest in the Novice class and lowest in the Expert class. The relatively lower F1 score and wider confidence interval for the Expert class may be partly explained by its underrepresentation in the dataset (n=45), which is substantially smaller than the other classes.

\begin{table*}[t]
\centering
\small
\setlength{\tabcolsep}{8pt}
\renewcommand{\arraystretch}{1.05}
\resizebox{0.95\textwidth}{!}{%
\begin{tabular}{lccccc}
\toprule
\textbf{Metric} & \textbf{Novice} ($n$=651) & \textbf{Beginner} ($n$=245) & \textbf{Intermediate} ($n$=209) & \textbf{Proficient} ($n$=120) & \textbf{Expert} ($n$=45) \\
\midrule
F1      & 0.8558 & 0.5365 & 0.6186 & 0.6114 & 0.4286 \\
95\% CI & [0.8349, 0.8757] & [0.4797, 0.5861] & [0.5669, 0.6667] & [0.5366, 0.6845] & [0.2727, 0.5683] \\
\bottomrule
\end{tabular}%
}
\caption{Per-class F1 scores with 95\% bootstrap confidence intervals from 1,000 resamples with replacement. Overall macro-F1 is 0.6102 with 95\% CI [0.5682, 0.6479].}
\label{tab:per-class-f1}
\end{table*}
}
\rev{\section{In-depth Statistical Analyses of User Keystroke Behavior}

\subsection{Expertise-level Differences}
\label{appendix:indepthkey_1}

\begin{table}[h]
\color{revcolor}
\captionsetup{labelfont={color=revcolor}, textfont={color=revcolor}}
\centering
\small
\caption{Kruskal-Wallis test results for expertise-level differences in keystroke-based behavioral features. Keystroke timing features are aggregated at the word level (same as model input feature), yielding both mean and standard deviation statistics, while typing speed and backspace count are represented by single word-level values.}
\label{tab:kw_expertise}
\resizebox{\columnwidth}{!}{%
\begin{tabular}{lcccc}
\toprule
\multirow{2}{*}{\textbf{Metric}} 
& \multicolumn{2}{c}{\textbf{$p$-value}} 
& \multicolumn{2}{c}{\textbf{H-statistic}} \\
\cmidrule(lr){2-3} \cmidrule(lr){4-5}
& \textbf{Mean} & \textbf{Std} 
& \textbf{Mean} & \textbf{Std} \\
\midrule
Key Duration    
& $<0.001$ & 68.42 & $<0.001$  & 37.10 \\
DD time         
& $<0.001$ & 263.45 & $<0.001$ & 101.42 \\
UD time         
& $<0.001$ & 271.18 & $<0.001$ & 69.30 \\
UU time         
& $<0.001$ & 207.17 & $<0.001$ & 60.28 \\
DU time         
& $<0.001$ & 116.50 & $<0.001$ & 82.16 \\
\midrule
Typing Speed    
& \multicolumn{2}{c}{$<0.001$} 
& \multicolumn{2}{c}{262.84} \\
Backspace Count 
& \multicolumn{2}{c}{$<0.001$} 
& \multicolumn{2}{c}{58.79} \\
\bottomrule
\end{tabular}%
}
\end{table}

To assess whether keystroke-based behavioral features differ statistically across expertise levels, we conducted a Kruskal-Wallis test~\cite{Kruskal01121952}, a non-parametric method suitable for comparing multiple groups with potentially unequal sample sizes and non-normal distributions. To ensure consistency with the features used by \system and to mitigate dependence among raw keystroke events, we performed this analysis on word-level aggregated behavioral features, treating each typed word as a single behavioral unit. The analysis was conducted across all expertise levels defined in our study.

We examined a range of keystroke metrics, including key duration, Down-Down (DD), Up-Down (UD), Up-Up (UU), and Down-Up (DU) inter-key intervals, as well as typing speed and backspace count. The results show robust and highly significant differences across expertise groups for nearly all behavioral features (all $p < 0.001$). Corresponding H-statistics indicate clear separation between expertise-level distributions, with larger values reflecting stronger group-wise differentiation.

These results provide statistical evidence that keystroke dynamics systematically vary with user expertise. In particular, higher expertise levels are associated with faster, more stable typing patterns and fewer correction behaviors, consistent with increased fluency and conceptual familiarity. This expertise-aligned structure of behavioral signals supports our claim that behavioral cues complement semantic information by mitigating ambiguity in query text and enabling more accurate expertise inference.

\subsection{User-level Differences}
\label{appendix:indepthkey_2}

\begin{table}[h]
\color{revcolor}
\captionsetup{labelfont={color=revcolor}, textfont={color=revcolor}}
\centering
\caption{One-way ANOVA results for user-level differences in keystroke-based behavioral features using a calibration dataset.}
\label{tab:anova_user}
\resizebox{0.7\columnwidth}{!}{%
\setlength{\tabcolsep}{6pt}
\renewcommand{\arraystretch}{1.2}
\begin{tabular}{lcc}
\toprule
\textbf{Metric} & \textbf{$p$-value} & \textbf{F-statistic} \\
\midrule
Key Duration & $< 0.001$ & 42.06  \\
DD time      & $0.006$ & 1.66 \\
UD time      & $0.011$ & 1.59  \\
UU time      & $0.006$ & 1.66  \\
DU time      & $0.002$ & 1.77  \\
\bottomrule
\end{tabular}
}
\end{table}

To further investigate individual variability in keystroke behavior, we analyzed whether keystroke features differ significantly across users. We conducted a one-way ANOVA using a calibration dataset in which each participant typed a standardized paragraph before the main study, allowing for controlled comparison of typing behavior independent of task content.

The analysis included multiple keystroke features, such as key duration and DD, UD, UU, and DU inter-key intervals. The results reveal significant between-user differences for all examined features (all $p < 0.05$), with F-statistics indicating substantial separation among users. These findings confirm that keystroke behaviors are highly user-specific, consistent with prior studies on typing dynamics~\cite{keyr4,keyr5,keyr6}.

This strong user-level variability explains our empirical observation that within-user few-shot inference consistently outperforms cross-user settings. Because each user’s typing rhythm and fluency form statistically distinct distributions, models trained across users tend to average away meaningful individual differences, leading to degraded performance. In contrast, within-user examples preserve user-specific behavioral patterns, enabling more effective personalization.

\section{Full Examples of Model Predicted Expertise and Reasoning}
\label{appendix:expertise_inference_qual}
This appendix provides full qualitative examples illustrating how ~\system{} predicts user expertise and reasons over semantic and behavioral cues at the query level (Table~\ref{tab:case_study_full}. For each example, we present the original user input, the predicted expertise level, and the corresponding reasoning process that integrates semantic signals from the query text with behavioral signals derived from keystroke interactions. These examples complement the quantitative and statistical analyses in the main paper by offering concrete, case-level evidence of how ~\system{} adapts its inference to different users and contexts. Together, they demonstrate how the proposed framework resolves semantic ambiguity, leverages user-specific behavioral patterns, and produces expertise-aware predictions in practice.

\begin{table*}[t]
\color{revcolor}
\captionsetup{labelfont={color=revcolor}, textfont={color=revcolor}}
    \centering
    \small
    \caption{A full qualitative case examples illustrating \system{}’s semantic-behavioral integration.}
    \renewcommand{\arraystretch}{1.1}
    
    \begin{tabularx}{0.98\textwidth}{
      p{1.4cm}
      >{\raggedright\arraybackslash}X
      >{\raggedright\arraybackslash}X
    }
        \toprule
         Query & Tell me everything that you know about priority scheduling and multilevel queue scheduling  & Can you explain in detail correlation with spectroscopic properties(e.g. IR and Raman activity)? \\
        \midrule
        User-reported & {\centering\textbf{Expert}} & {\centering\textbf{Beginner}} \\
        \midrule
        \textbf{~\system{} (ours)}
        & \textbf{Prediction:} Expert. \newline \textbf{Reason:} The query requests ""everything you know about priority scheduling and multilevel queue scheduling,"" which is a broad, open-ended prompt typical of users with high confidence and deep interest in comprehensive coverage. The user uses precise technical terms (""priority scheduling,"" ""multilevel queue scheduling"") without hesitation or correction in those segments, as shown by stable keystroke metrics (low backspace, consistent timing, high typing speed). This matches the pattern in Example 7, where similar open-ended, technical queries and fluent typing are associated with Expert-level users. There is no evidence of definition-seeking or conceptual uncertainty. The baseline for this user, given the query style and typing, is Expert, and no DELTA adjustment is warranted.
        & \textbf{Prediction:} Beginner. \newline \textbf{Reason:} Baseline from prior queries (see Example 10) is Beginner, with consistent requests for careful explanation and explicit self-identification as a beginner. The current query uses technical terms (""correlation with spectroscopic properties"", ""IR"", ""Raman activity"") but only in the context of asking for detailed explanation, not applying or linking concepts. Typing behavior on technical segments is stable (no strong hesitation or correction), so no negative adjustment is warranted, but there is also no evidence for higher conceptual mastery or application. Final level remains Beginner.\\
        \midrule
        Semantic-only
        & \textbf{Prediction:} Novice. \newline \textbf{Reason:} he query is broad and non-specific, lacking any technical terminology or focused questions. The user requests ""everything"" about the topics, which is characteristic of someone with little or no familiarity seeking a general overview.
        & \textbf{Prediction:} Beginner. \newline \textbf{Reason:} The user requests a detailed explanation of the correlation between spectroscopic properties and provides examples (IR and Raman activity), but does not use specialized terminology or demonstrate deeper understanding of the underlying concepts. The phrasing suggests curiosity and a need for foundational explanation, consistent with a beginner level. \\
        \midrule
        Behavior-only
        & \textbf{Prediction:} Proficient. \newline \textbf{Reason:} Typing speed is consistently high (most words above 300 kpm, several above 700 kpm), with low to moderate backspace usage (only 3 total across 12 words), and moderate variability in timing features. The user demonstrates fluent, confident typing with low error correction, and the timing metrics (mean and std) are within the range of proficient or expert users in the examples. However, the variability in some timing features (e.g., UD-time std up to 77.60, 133.84, 223.49) is higher than typical expert patterns, and the user does not show the extremely low error rates or the highly consistent timing seen in expert examples. This places the user solidly in the proficient category, with strong fluency but not the hallmark consistency of an expert.
        & \textbf{Prediction:} Proficient. \newline \textbf{Reason:} Typing speed is consistently high (276–983 keystrokes per minute), with low to moderate backspace usage (zero throughout), and moderate variability in timing features. The user demonstrates fluent, confident typing with low error correction, and the timing metrics (mean and std) are stable and similar to those seen in proficient or expert examples. However, the user does not show the extremely low variability or the highly consistent, advanced timing patterns typical of experts in the provided examples. The overall pattern is more advanced than intermediate but does not reach the expert level. \\
        \bottomrule
s    \end{tabularx}
    \label{tab:case_study_full}
\end{table*}

\section{Qualitative Comparison of Baseline and ~\system{} Responses}
\label{apx:~\system{} Responses}
Tables~\ref{tab:qualitative_novice} to~\ref{tab:qualitative_expert} show examples of baseline and ~\system{} responses at different expertise levels.

\begin{table*}[h]
\color{revcolor}
\captionsetup{labelfont={color=revcolor}, textfont={color=revcolor}}
\centering
\small
\renewcommand{\arraystretch}{1.4}
\caption{Qualitative examples comparing baseline and expertise-conditioned (~\system{}) responses in Novice.}
\begin{tabular}{p{0.15\linewidth} p{0.75\linewidth}}
\hline
\multicolumn{2}{l}{\textbf{Example 1 — Chemistry Query (Expertise: Novice, P10):}} \\
\hline
User Query &
I do not understand what “molecular symmetry” means. Please explain it in a simple and intuitive way, using easy examples, as if you were explaining it to an elementary school student. \\
Without Expertise \newline (baseline) &
“Of course. I will explain “molecular symmetry” in a very simple way, using familiar and easy examples that even an elementary school student can understand.
First, think about a mirror. When you place a mirror in front of your face, you see the same face reflected back. This happens because your face is reflected symmetrically in the mirror. In general, symmetry means that when something is divided into two parts, the two sides look the same or very similar. For example, a butterfly’s wings look almost identical on the left and right sides, so we say that a butterfly has symmetry.
Now, what does “molecular symmetry” mean? A molecule is an extremely small particle. The air we breathe, the water we drink, and the food we eat are all made of molecules. Molecules also have shapes. Some molecules, like a butterfly, look the same on the left and right sides. Other molecules look the same even when you rotate them.
For example, a water molecule (H₂O) is made of one oxygen atom and two hydrogen atoms. The hydrogen atoms are attached at an angle, so the shape is slightly bent, and the symmetry is not perfect. In contrast, a carbon dioxide molecule (CO₂) has a carbon atom in the center and identical oxygen atoms on both sides, which makes it highly symmetric.
In summary, molecular symmetry refers to whether the shape of a molecule looks the same when it is reflected in a mirror or rotated. This concept is important because scientists use molecular symmetry to understand and predict the properties of molecules.” \\
With Expertise \newline (~\system{}, ours) &
“I will explain “molecular symmetry” in a very simple way.
First, a molecule is a small particle that makes up matter. For example, even a single drop of water contains an enormous number of water molecules. A water molecule is called H₂O, meaning it is made of two hydrogen atoms and one oxygen atom.
Next, let us think about what symmetry means. Symmetry refers to a situation in which an object looks the same on both sides when it is divided in half. Common examples include butterfly wings, human faces (approximately), and heart shapes. When the right and left sides look the same, we say that the object is symmetric.
Molecules are very small, but they also have shapes. If a molecule looks the same when you rotate it or reflect it, we say that it has molecular symmetry.
For example, a water molecule (H₂O) has two hydrogen atoms attached to a central oxygen atom, but the shape is bent, which slightly breaks the symmetry. On the other hand, a carbon dioxide molecule (CO₂) has the structure O = C = O, with identical oxygen atoms on both sides of the carbon atom. Because of this, its shape remains the same when viewed from either side, making it a highly symmetric molecule.
In simple terms, molecular symmetry means checking whether a molecule looks the same when it is rotated or flipped. This symmetry can influence the properties of a molecule, such as how it interacts with light or how it reacts with other substances.” \\
\hline
\end{tabular}
\label{tab:qualitative_novice}
\end{table*}

As shown in Table \ref{tab:qualitative_novice}, the ~\system{} response demonstrates a deliberate restructuring of the explanation to match the cognitive constraints of novice users. Rather than merely simplifying terminology, ~\system{} decomposes the target concept into a sequence of prerequisite notions (e.g., molecule → symmetry → molecular symmetry), thereby explicitly scaffolding understanding. This layered progression contrasts with the baseline response, which presents a coherent but largely narrative explanation without clearly isolating conceptual dependencies. The insight here is that ~\system{} does not only simplify content for novices, but actively reorganizes knowledge to minimize abstraction and reduce cognitive load, enabling comprehension without prior domain assumptions.

\begin{table*}[h]
\color{revcolor}
\captionsetup{labelfont={color=revcolor}, textfont={color=revcolor}}
\centering
\small
\renewcommand{\arraystretch}{1.4}
\caption{Qualitative examples comparing baseline and expertise-conditioned (~\system{}) responses in Beginner.}
\begin{tabular}{p{0.15\linewidth} p{0.75\linewidth}}
\hline
\multicolumn{2}{l}{\textbf{Example 2 — Business Query (Expertise: Beginner, P15):}} \\
\hline
User Query &
So, for example, if a parent company invests 10 billion won in a subsidiary and the subsidiary's net assets are 8 billion won, then a 2 billion won operating right will be generated. Will this 2 billion won be recorded in the subsidiary's accounting books? \\
Without Expertise \newline (baseline) &
“This is a good question. Below is an explanation of goodwill arising in the relationship between a parent company and a subsidiary under accounting principles. In summary, goodwill of KRW 2 billion is not recorded in the subsidiary’s accounting books (\xmark), but is recognized only in the parent company’s consolidated financial statements (\cmark). To explain why, when a parent company acquires a subsidiary for KRW 10 billion while the subsidiary’s net assets (assets minus liabilities) amount to KRW 8 billion, the parent company pays more than the book value of the subsidiary’s net assets. The difference of KRW 2 billion represents goodwill, which reflects intangible values such as brand reputation, customer relationships, and know-how. However, this goodwill is not included in the subsidiary’s own accounting records, as the subsidiary records only its own assets, liabilities, and equity. The goodwill of KRW 2 billion is recognized as an asset only when the parent company prepares consolidated financial statements and accounts for the acquisition of the subsidiary. In conclusion, goodwill does not appear in the subsidiary’s books (\xmark), is recorded as an equity investment in the parent company’s separate financial statements (\cmark), and is recognized as a goodwill asset of KRW 2 billion in the parent company’s consolidated financial statements (\cmark)." \\
With Expertise \newline (~\system{}, ours) &
“This is a good question. The key concept here is the distinction between consolidated financial statements and separate financial statements. Consider the following situation: a parent company acquires the shares of a subsidiary for KRW 10 billion, and the subsidiary’s net assets amount to KRW 8 billion, resulting in a difference of KRW 2 billion that corresponds to goodwill. The core question is whether this KRW 2 billion of goodwill is recorded in the subsidiary’s accounting books. The answer is no. Goodwill is not recorded in the subsidiary’s books. The subsidiary prepares its accounting records from its own perspective, recognizing only its own assets and liabilities, and the price paid by an external investor to acquire its shares is not relevant from the subsidiary’s standpoint. In contrast, the parent company pays an amount exceeding the subsidiary’s net assets at the time of acquisition, and this excess amount, calculated as KRW 10 billion minus KRW 8 billion, is recognized as goodwill in the parent company’s consolidated financial statements. In simple terms, goodwill is not recorded in the subsidiary’s separate financial statements (\xmark) or in the parent company’s separate financial statements (\xmark), but is recognized as a KRW 2 billion goodwill asset in the consolidated financial statements of the parent and subsidiary (\cmark), which are prepared as if the two entities were a single economic entity.” \\
\hline
\end{tabular}

\label{tab:qualitative_beginner}
\end{table*}

As shown in Table \ref{tab:qualitative_beginner}, the ~\system{} response shifts the explanatory focus from outcome reporting to perspective-oriented reasoning. Specifically, the ~\system{} response foregrounds the distinction between separate and consolidated financial statements before addressing the treatment of goodwill, making the underlying accounting logic explicit. By contrast, the baseline response primarily states where goodwill is recorded, offering limited insight into why that treatment arises. This indicates that, at the beginner level, ~\system{} emphasizes conceptual framing and viewpoint alignment, helping users internalize domain-specific reasoning patterns rather than memorizing isolated results.

\begin{table*}[h]
\color{revcolor}
\captionsetup{labelfont={color=revcolor}, textfont={color=revcolor}}
\centering
\small
\renewcommand{\arraystretch}{1.4}
\caption{Qualitative examples comparing baseline and expertise-conditioned (~\system{}) responses in Intermediate.}
\begin{tabular}{p{0.15\linewidth} p{0.75\linewidth}}
\hline
\multicolumn{2}{l}{\textbf{Example 3 — Computer Science Query (Expertise: Intermediate, P10):}} \\
\hline
User Query &
In operating systems, it seems that Round Robin scheduling is used not only in FCFS-based scheduling but also within priority scheduling and multilevel queue scheduling. Please explain this clearly with simple examples. \\
Without Expertise \newline (baseline) &
“Priority scheduling and multilevel queue scheduling are representative scheduling algorithms used in operating systems to efficiently manage multiple processes. In both approaches, processes are organized into queues, and different scheduling policies can be applied within each queue. In addition to First-Come, First-Served (FCFS), Round Robin (RR) scheduling may also be used. Round Robin assigns a fixed time quantum to each process and executes processes in a cyclic order.
Consider the following example. Suppose there are three processes in the system: P1 with high priority, P2 with medium priority, and P3 with low priority. Each priority level corresponds to a separate queue, and each queue uses a different scheduling policy. The highest-priority queue uses Round Robin with a time quantum of 2 ms, the medium-priority queue uses FCFS, and the lowest-priority queue uses Round Robin with a time quantum of 4 ms.
Assume that P1 arrives at time 0 ms with a burst time of 5 ms in the highest-priority queue, P2 arrives at time 1 ms with a burst time of 4 ms in the medium-priority queue, and P3 arrives at time 2 ms with a burst time of 8 ms in the lowest-priority queue. Since the highest-priority queue is always selected first, P1 executes using Round Robin until completion. Only after the highest-priority queue becomes empty does P2 execute using FCFS, followed by P3 executing with Round Robin in the lowest-priority queue.
This example illustrates that Round Robin operates only within individual queues, while the selection among queues is determined by priority. In multilevel queue scheduling, different scheduling policies such as FCFS or Round Robin can be applied independently within each queue, depending on the characteristics of the processes assigned to that queue.” \\
With Expertise \newline (~\system{}, ours) &
“Priority scheduling and multilevel queue scheduling operate differently, but both can incorporate Round Robin (RR) scheduling as an internal mechanism. Round Robin assigns each process an equal time slice, allowing processes to share the CPU in a fair and cyclic manner.
In multilevel queue scheduling, processes are classified into separate queues based on their characteristics, and each queue may use a different scheduling algorithm. For example, a system may use FCFS for real-time processes, Round Robin for interactive processes, and FCFS or Shortest Job First for background batch jobs. Within a queue that uses Round Robin, processes take turns executing for a fixed time quantum, while the scheduling order among queues is determined by their priority.
Similarly, in priority scheduling, the CPU is allocated to processes with higher priority first. However, when multiple processes share the same priority level, Round Robin scheduling can be applied among those processes to ensure fairness. For instance, if several processes have identical priority values, they can be scheduled in a Round Robin fashion with a fixed time slice.
In summary, Round Robin is commonly used to provide fairness either within a specific queue in multilevel queue scheduling or among processes with the same priority in priority scheduling. This combination allows operating systems to balance responsiveness and efficiency while respecting priority constraints.” \\
\hline
\end{tabular}
\label{tab:qualitative_intermediate}
\end{table*}

As shown in Table \ref{tab:qualitative_intermediate}, the ~\system{} response transitions from procedural illustration to rule-based abstraction. While the baseline response relies on a detailed scheduling scenario to demonstrate how Round Robin operates within priority and multilevel queue scheduling, ~\system{} extracts general principles (e.g., Round Robin as an intra-queue fairness mechanism or as a tie-breaking strategy among equal-priority processes). The key insight is that ~\system{} adapts to intermediate users by prioritizing transferable reasoning: explanations are structured so that the user can apply the same logic to unseen system configurations, rather than reproducing a single execution trace.

\begin{table*}[h]
\color{revcolor}
\captionsetup{labelfont={color=revcolor}, textfont={color=revcolor}}
\centering
\small
\renewcommand{\arraystretch}{1.4}
\caption{Qualitative examples comparing baseline and expertise-conditioned (~\system{}) responses in Proficient.}
\begin{tabular}{p{0.15\linewidth} p{0.75\linewidth}}
\hline
\multicolumn{2}{l}{\textbf{Example 4 — Computer Science Query (Expertise: Proficient, P17):}} \\
\hline
User Query &
Give examples of priority scheduling and multilevel queue scheduling. \\
Without Expertise \newline (baseline) &
“Of course. First, I will briefly explain each scheduling method and then provide corresponding examples. 1. Priority Scheduling Concept: Priority scheduling assigns a priority to each process and allocates the CPU to the process with the highest priority first. In many systems, a smaller numerical value indicates a higher priority. Example: Scenario: Assume the following four processes exist in the system. Process P1 arrives at time 0 with a burst time of 5 and priority 2; P2 arrives at time 1 with a burst time of 3 and priority 1; P3 arrives at time 2 with a burst time of 8 and priority 4; and P4 arrives at time 3 with a burst time of 6 and priority 3. A lower numerical priority value indicates a higher priority. If only priority values were considered, the order would be P2, P1, P4, and P3. However, when arrival time is also considered, the scheduler selects the highest-priority process among those that have already arrived. The actual scheduling order under non-preemptive priority scheduling is as follows: at time 0, P1 starts executing; at time 5, P2 executes since it has the highest priority among the arrived processes; at time 8, P4 executes; and at time 14, P3 executes. Note that in the preemptive version of priority scheduling, if a process with a higher priority arrives later, it can interrupt the currently running process and take over the CPU. 2. Multilevel Queue Scheduling Concept: Multilevel queue scheduling classifies processes into multiple queues based on their characteristics, assigns different scheduling algorithms to each queue, and may also define priorities among the queues themselves. Example: Suppose the system has three queues. Queue 1 contains system processes with the highest priority and uses First-Come, First-Served scheduling. Queue 2 contains interactive processes and uses Round Robin scheduling with a time quantum of 4. Queue 3 contains batch jobs and uses First-Come, First-Served scheduling. Assume Queue 1 contains process P1 with a burst time of 3, Queue 2 contains process P2 with a burst time of 10, and Queue 3 contains process P3 with a burst time of 15. With inter-queue priorities applied, P1 executes first, followed by P2, and finally P3. The execution proceeds as follows: from time 0 to 3, P1 completes execution; from time 3 to 7, P2 executes its first time slice, leaving 6 units of remaining time; from time 7 to 11, P2 executes its second time slice, leaving 2 units; from time 11 to 13, P2 executes its third time slice and completes; from time 13 to 28, P3 executes. Summary of differences: Priority scheduling determines execution order based on per-process priority values within a single queue and may be preemptive or non-preemptive, whereas multilevel queue scheduling uses multiple queues organized by process type, with different scheduling algorithms and priorities assigned to each queue.” \\
With Expertise \newline (~\system{}, ours) &
“Certainly. Below is an explanation of priority scheduling and multilevel queue scheduling, along with illustrative examples. 1. Priority Scheduling Concept: Priority scheduling assigns a priority to each process, and the CPU executes the process with the highest priority first. Depending on the system, a higher priority may be represented by a smaller or larger numerical value. Priorities may be static or dynamic. In non-preemptive priority scheduling, a running process retains the CPU until it finishes, whereas in preemptive priority scheduling, the arrival of a higher-priority process causes the currently running process to be suspended. Example: Consider three processes. Process P1 arrives at time 0 ms with a burst time of 5 ms and priority 2; process P2 arrives at time 1 ms with a burst time of 3 ms and priority 1; process P3 arrives at time 2 ms with a burst time of 1 ms and priority 3, where a smaller number indicates a higher priority. Under preemptive priority scheduling, P1 begins execution at time 0. At time 1, P2 arrives with a higher priority than P1, so P1 is preempted and P2 executes. At time 4, P2 completes and P1 resumes execution. At time 9, P1 completes and P3 executes. 2. Multilevel Queue Scheduling Concept: Multilevel queue scheduling divides processes into multiple categories, such as interactive processes, user applications, and background tasks, and applies an independent scheduling algorithm to each queue. Priorities also exist among the queues, such that a lower-priority queue is executed only when all higher-priority queues are empty. Representative queue classifications include Queue 0 for interactive processes using Round Robin scheduling, Queue 1 for user processes using Shortest Job First scheduling, and Queue 2 for background applications using First-Come, First-Served scheduling. Example: Process P1 is an interactive user process in Queue 0 with Round Robin scheduling and a time quantum of 2 ms; P2 is a user application in Queue 1 using Shortest Job First; and P3 is a background backup task in Queue 2 using First-Come, First-Served scheduling. In execution, P1 in Queue 0 runs first for 2 ms. Once Queue 0 becomes empty, the scheduler moves to Queue 1 and executes P2. Only after Queue 1 finishes does P3 execute, since Queue 2 has the lowest priority. In summary, priority scheduling determines execution order by assigning explicit priorities to individual processes, whereas multilevel queue scheduling separates processes by type and assigns different algorithms and priorities to each queue.” \\
\hline
\end{tabular}
\label{tab:qualitative_proficient}
\end{table*}

As shown in Table \ref{tab:qualitative_proficient}, the ~\system{} response emphasizes conceptual taxonomy and awareness of the design space. Instead of extending procedural examples, ~\system{} explicitly categorizes scheduling policies along orthogonal dimensions such as static versus dynamic priorities and preemptive versus non-preemptive execution. The baseline response, although comprehensive, remains example-heavy and less explicit about these structural distinctions. This suggests that ~\system{} aligns with proficient users’ expectations by highlighting comparative frameworks and architectural trade-offs, supporting analytical reasoning rather than operational understanding alone.

\begin{table*}[h]
\color{revcolor}
\captionsetup{labelfont={color=revcolor}, textfont={color=revcolor}}
\centering
\small
\renewcommand{\arraystretch}{1.4}
\caption{Qualitative examples comparing baseline and expertise-conditioned (~\system{}) responses in Expert.}
\begin{tabular}{p{0.15\linewidth} p{0.75\linewidth}}
\hline
\multicolumn{2}{l}{\textbf{Example 5 — Chemistry Query (Expertise: Expert, P23):}} \\
\hline
User Query &
When classifying point groups based on molecular symmetry, is it correct that infrared (IR) spectroscopy probes changes in the dipole moment, while Raman spectroscopy probes changes in polarizability, and that peaks are observed at the corresponding vibrational frequencies that satisfy these conditions? \\
Without Expertise \newline (baseline) &
“Yes, this understanding is correct. When molecular symmetry and point group theory are applied, it becomes possible to predict which vibrational modes of a molecule are observable using different spectroscopic techniques such as infrared (IR) and Raman spectroscopy.
Infrared spectroscopy detects vibrational modes that induce a change in the molecular dipole moment, denoted as $\Delta \mu$. For a vibrational mode to be IR-active, the vibration must produce a time-dependent change in the dipole moment of the molecule. Typical examples include asymmetric stretching and asymmetric bending modes.
Raman spectroscopy, in contrast, detects vibrational modes that cause a change in the molecular polarizability, denoted as $\Delta \alpha$. In this case, the key factor is how the electronic cloud of the molecule is distorted during vibration. A vibrational mode is Raman-active only if it produces a modulation of the polarizability. Symmetric stretching modes are common examples of Raman-active vibrations.
As a result, peaks observed at specific vibrational frequencies in IR and Raman spectra arise from different physical activation conditions. By using point group theory and character tables, one can determine which vibrational modes are IR-active and which are Raman-active, allowing theoretical predictions to be directly compared with experimental observations.
For example, carbon dioxide (CO$2$) is a linear molecule belonging to the $D{\infty h}$ point group. Some of its vibrational modes are IR-active but Raman-inactive, while others are Raman-active but IR-inactive. This complementarity explains why IR and Raman spectroscopy are often used together in vibrational analysis.
In summary, IR activity requires $\Delta \mu \neq 0$ during vibration, while Raman activity requires $\Delta \alpha \neq 0$.” \\
With Expertise \newline (~\system{}, ours) &
“Yes, your understanding is accurate. Infrared and Raman spectroscopy probe different vibrational modes depending on molecular symmetry, and this distinction is fundamentally based on changes in physical properties, namely the dipole moment and the polarizability.
Infrared spectroscopy measures time-dependent changes in the dipole moment during molecular vibration. A normal mode is IR-active if it produces a nonzero change in the dipole moment during vibration. Consequently, vibrational modes that are asymmetric with respect to the molecular geometry are most commonly observed in IR spectra. For instance, although carbon dioxide (CO$_2$) is a linear and overall symmetric molecule, its asymmetric stretching and bending modes generate dipole moment changes and therefore appear in the IR spectrum.
Raman spectroscopy probes changes in the molecular polarizability, more precisely changes in the polarizability tensor. A vibrational mode is Raman-active if the vibration causes a modulation of the polarizability. As a result, symmetric vibrational modes often give strong Raman signals. In the case of CO$_2$, the symmetric stretching mode does not induce a dipole moment change and is therefore IR-inactive, but it does change the polarizability and thus appears prominently in the Raman spectrum.
From the symmetry analysis perspective, once the molecular point group is identified, each vibrational mode can be assigned to an irreducible representation. Selection rules derived from the character table determine whether a given mode is IR-active, corresponding to the Cartesian coordinates $x$, $y$, or $z$, or Raman-active, corresponding to quadratic functions such as $x^2$, $y^2$, $z^2$, or $xy$.
In summary, IR spectroscopy is governed by the condition $\Delta \mu \neq 0$, whereas Raman spectroscopy is governed by the condition $\Delta \alpha \neq 0$. The two techniques, therefore, provide complementary information for the analysis of molecular vibrational modes.” \\
\hline
\end{tabular}
\label{tab:qualitative_expert}
\end{table*}}

\color{revcolor}
\captionsetup{labelfont={color=revcolor}, textfont={color=revcolor}}
As shown in Table \ref{tab:qualitative_expert}, the ~\system{} response demonstrates expert-level adaptation by embedding explanations within formal theoretical structures. The ~\system{} response connects IR and Raman activity to group-theoretic constructs such as irreducible representations, polarizability tensors, and selection rules derived from character tables. In contrast, the baseline response remains descriptive, focusing on physical conditions without explicitly linking them to symmetry formalism. The insight here is that ~\system{} does not merely add detail, but shifts the explanatory goal toward theoretical completeness and disciplinary rigor, reflecting how experts evaluate correctness and depth through formal frameworks rather than phenomenological description.

\section{License for Artifacts}
We follow the terms of use released by OpenAI (\url{https://openai.com/policies/row-terms-of-use/}).

\section{Potential Risks}
The proposed framework dynamically adapts large language model responses based on inferred user expertise. While this personalization is designed to improve user experience, it introduces potential societal risks. Expertise misclassifications could systematically disadvantage particular user groups or amplify pre-existing biases, especially for users with non-standard interaction patterns. Additionally, malicious actors could attempt to manipulate the system to receive inappropriate or misleading personalized outputs. 

\section{Implementation Details and Evaluation Packages}

We used standard Python libraries for evaluation and statistical analysis.
Evaluation metrics, including mean absolute error (MAE), mean squared error (MSE), and confusion matrices, were computed using the \texttt{sklearn.metrics} package with default parameter settings.
Statistical significance testing was conducted using standard functions from the \texttt{scipy.stats} package.
Specifically, the Wilcoxon signed-rank test using \texttt{scipy.stats.wilcoxon}, one-way ANOVA was performed using \texttt{scipy.stats.f\_oneway}, and the Kruskal--Wallis test using \texttt{scipy.stats.kruskal}.
All tests were applied following their conventional assumptions and default configurations.

\end{document}